\documentclass[
 aps,
 pra,
 showpacs,
 showkeys,
 superscriptaddress,
 reprint,%
twocolumn,
]{revtex4-1}
\usepackage{physics}
\usepackage{mathtools}
\usepackage{amssymb, amsfonts, amsthm}
\usepackage{siunitx}
\usepackage[toc,page]{appendix}
\usepackage[caption=false]{subfig}
\usepackage{bm}
\usepackage{soul}
\usepackage[dvipsnames]{xcolor}

\usepackage{hyperref}
\hypersetup{colorlinks=true, linkcolor=blue, citecolor=blue, urlcolor=blue, unicode=true}
\usepackage{cleveref}

\begin{document}
\title{Exploring entanglement and optimization within the Hamiltonian Variational Ansatz}

\author{Roeland Wiersema}
\email{rwiersema@uwaterloo.ca}
\affiliation{Vector Institute, MaRS  Centre,  Toronto,  Ontario,  M5G  1M1,  Canada}
\affiliation{Department of Physics and Astronomy, University of Waterloo, Ontario, N2L 3G1, Canada}

\author{Cunlu Zhou}
\email{czhou@cs.toronto.edu}
\affiliation{Department of Computer Science, University of Toronto, Ontario, M5T 3A1, Canada}
\affiliation{Vector Institute, MaRS  Centre,  Toronto,  Ontario,  M5G  1M1,  Canada}

\author{Yvette de Sereville}
\affiliation{Department of Physics and Astronomy, University of Waterloo, Ontario, N2L 3G1, Canada}
\affiliation{Institute for Quantum Computing, University of Waterloo, Ontario, N2L 3G1, Canada}

\author{Juan Felipe Carrasquilla}
\affiliation{Vector Institute, MaRS  Centre,  Toronto,  Ontario,  M5G  1M1,  Canada}
\affiliation{Department of Physics and Astronomy, University of Waterloo, Ontario, N2L 3G1, Canada}

\author{Yong Baek Kim}
\affiliation{Department of Physics, University of Toronto, Ontario M5S 1A7, Canada}

\author{Henry Yuen}
\affiliation{Department of Computer Science and Department of Mathematics, University of Toronto, Ontario, M5T 3A1, Canada}

\date{\today}

\begin{abstract}

    Quantum variational algorithms are one of the most promising applications of near-term quantum computers; however, recent studies have demonstrated that unless the variational quantum circuits are configured in a problem-specific manner, optimization of such circuits will most likely fail. In this paper, we focus on a special family of quantum circuits called the Hamiltonian Variational Ansatz (HVA), which takes inspiration from the quantum approximation optimization algorithm and adiabatic quantum computation. Through the study of its entanglement spectrum and energy gradient statistics, we find that HVA exhibits favorable structural properties such as mild or entirely absent barren plateaus and a restricted state space that eases their optimization in comparison to the well-studied ``hardware-efficient ansatz.'' We also numerically observe that the optimization landscape of HVA becomes almost trap free, i.e., no sub-optimal minima, when the ansatz is over-parameterized. We observe a size-dependent ``computational phase transition'' as the number of layers in the HVA circuit is increased where the optimization crosses over from a hard to an easy region in terms of the quality of the approximations and speed of convergence to a good solution. In contrast with the analogous transitions observed in the learning of random unitaries which occur at a number of layers that grows exponentially with the number of qubits, our Variational Quantum Eigensolver experiments suggest that the threshold to achieve the over-parameterization phenomenon scales at most polynomially in the number of qubits for the transverse field Ising and XXZ models. Lastly, as a demonstration of its entangling power and effectiveness, we show that HVA can find accurate approximations to the ground states of a modified Haldane-Shastry Hamiltonian on a ring, which has long-range interactions and has a power-law entanglement scaling.

\end{abstract}



\maketitle
\section{Introduction}

With the advent of Noisy Intermediate-Scale Quantum (NISQ) computers~\cite{Preskill2018nisq}, near-term quantum algorithms such as the Variational Quantum Eigensolvers (VQE), may offer computational capabilities beyond the best current classical computers and algorithms for approximating ground states of quantum many-body systems. A VQE algorithm contains three ingredients: A variational quantum circuit ansatz specified by a set of parameters $\bm{\theta}$, an energy function given by the expectation value of a local Hamiltonian $H$ composed of local measurements on the variational circuit state and a classical optimizer. A natural first approach is the random quantum circuit ansatz~\cite{Kandala2017hwefficient, Sim2019expressibility, bravoprieto2020scaling}, capable of expressing a wide variety of states. However, this was shown to be ineffective for gradient-based optimization strategies due to the barren plateau phenomenon~\cite{McClean2018barren, sharma2020trainability, volkoff2020large, cerezo2020costfunctiondependent}, which causes the optimization of randomly initialized circuits to get stuck on flat areas in the cost landscape where gradients are exponentially small. These observations suggest that an effective ansatz for VQE requires a circuit that is problem-specific, such that the optimization landscape of the problem is not hindered by barren plateaus.  For quantum many-body problems,  Ref.~\cite{Wecker2015} suggests a novel variational circuit that is now called the Hamiltonian Variational Ansatz (HVA). While there is no rigorous proof that HVA will be an effective ansatz, recent work has demonstrated that HVA is rather effective for several one- and two-dimensional quantum many-body models~\cite{Ho2019, Cade2019}. It is thus an intriguing question to further understand the empirically observed effectiveness of HVA.

For the purpose of understanding the effectiveness of such ans\"atze, it is useful to note that quantum entanglement provides a window into the capabilities of several families of numerical techniques and algorithms aimed at understanding the properties of quantum many-body states, as well as helps us delineate the boundary between quantum states that can be simulated classically and those which call for quantum simulators and quantum computers for their accurate description. For instance, for a one-dimensional (1D) gapped local Hamiltonian, the entanglement entropy of the ground state obeys an area law, i.e., the entanglement entropy grows proportional to the boundary area of the system instead of the system size~\cite{Eisert2010arealaw}. This remarkable result allows us to combat the exponential scaling of the Hilbert space, since this area law provides evidence that the relevant physics of a system only takes place in a restricted part of the full state space.  These observations have inspired a variety of variational numerical methods, most notably, Tensor Network approaches such as Matrix Product State (MPS), Multiscale Entanglement Renormalization and Projected Entangled Pair States~\cite{Eisert2013tensornetworks}, but also deep learning inspired variational approaches, which have been successful at representing quantum many-body states~\cite{Carleo2016nqs, Carleo2020autoreg,Carrasquilla2019transformers, HibatAllah2020rnns}.

In this paper we study various entanglement properties of HVA and present several results on the favorable features of HVA that shed light on the underlying reasons for its effectiveness for solving natural many-body problems. Our findings suggest that HVA is highly expressive but yet structured enough to allow for efficient optimization. Through the study of two prototypical models in condensed matter physics, namely the 1D transverse field Ising (TIFM) and XXZ models, we find that entanglement entropy and entanglement spectrum can shed light onto the initialization and optimization properties of HVA in the context of the VQE algorithm. Whereas HVA provides a restricted and effective state space for the TFIM which yields ground state approximations largely insensitive to the circuit initialization, the 1D XXZ model ansatz requires a careful parameter initialization for its successful optimization.  Through the study of the dynamics of entanglement spectrum during the optimization of the XXZ model we find that initializing the HVA near the identity operator enables a restricted and effective subspace during optimization that yields accurate approximations to the ground state with fast convergence. Furthermore, we show evidence that the gradient vanishing problem in HVA, especially if the HVA is initialized near the identity operator, is mild or entirely absent in comparison to the random circuit ansatz, where barren plateaus in the energy landscape cause gradients to decay exponentially with increasing system size. We also explore the over-parameterization phenomena in HVA and observe a ``computational phase transition'' between an under-parameterized and over-parameterized regime where the  optimization landscape of HVA crosses over to a regime with faster convergence and absence of low-quality solutions.  Lastly, as a demonstration of the entangling power and effectiveness of HVA, we study a modified Haldane-Shastry (MHS) Hamiltonian which has long-range interactions and a power-law scaling entanglement entropy~\cite{Deng2017}. We observe that HVA can find approximations to the ground state of the MHS Hamiltonian reaching fidelities $>99\%$ for system sizes $N=4,8,12,16$  and circuit depths $p=N$. Our findings point to important features of HVA that will lead to a deeper understanding of its effectiveness, and point the way to developing more sophisticated ans\"atze for other many-body problems, as well as more informed optimization strategies. Moreover,we establish a substantial connection between quantum entanglement and the efficacy of HVA and show how entanglement properties such as the entanglement spectrum can be used to study variational quantum circuits. Furthermore, the surprising phenomenon of over-parameterization in HVA signals a nontrivial connection with deep neural networks which merits further investigation.

In \cref{sec:vqehva} we introduce the basic concepts of VQE and HVA. In \cref{sec:methods}, we introduce two paradigmatic quantum many-body models which we use in our study, the Transverse Field Ising Model and the XXZ model, as well as their respective ans\"atze. We also introduce the necessary entanglement concepts used in this paper. In \cref{sec:main}, our main results are presented, and we conclude in \cref{sec:conclusion}. In the appendices, we include the computational details in \cref{app:compdet}, some additional numerical results in \cref{app:numres} and extra results on the dynamics of entanglement entropy in \cref{app:entropydynamics}.


\section{Variational Quantum Eigensolver and Hamiltonian Variational Ansatz}\label{sec:vqehva}

VQE~\cite{Peruzzo2014} is a hybrid classical-quantum algorithm for finding eigenstates of a quantum many-body Hamiltonian. According to the variational principle of quantum mechanics, a parameterized wave function $\ket{\psi(\bm{\theta})}$ provides an upper bound on the ground state energy, 
\begin{equation}
    E_{\text{ground}} \leq \bra{\psi(\bm{\theta})} H \ket{\psi(\bm{\theta})} = E(\bm{\theta}), \label{eq:cost}
\end{equation}
where $H$ is a $k$-local lattice Hamiltonian. Hence, we can approximate the ground state by minimizing $E(\bm{\theta})$ with respect to the parameters $\bm{\theta}$. In the case of VQE, the wave function $\ket{\psi(\bm{\theta})}$ corresponds to a depth $p$ quantum circuit specified by a unitary matrix $U(\bm{\theta})$, i.e., $\ket{\psi(\bm{\theta})} = U(\bm{\theta})\ket{0}$ where a number of $m$ parameters specify the unitary $\bm{\theta}\in \mathbb{R}^m$. We can estimate the variational energy $E(\bm{\theta})_p$, where $E(\bm{\theta})_{p}$ denotes the energy at $p$-level circuit, by measuring the observables that compose the Hamiltonian of the system over the quantum state $U(\bm{\theta})\ket{0}$. We use a classical optimization procedure to find the optimal parameters $\bm{\theta}^*$ that minimize the energy. 

As with other variational methods for approximating the ground state, a key ingredient to the success of the method is finding a good parameterization scheme of the wave function. Ideally, the manifold of states parameterized by the ansatz of choice contains the ground state of interest, and this ground state can be reached using a numerical optimization. The Hamiltonian Variational Ansatz (HVA)~\cite{Wecker2015} is a quantum circuit ansatz inspired by the quantum approximation optimization algorithm (QAOA)~\cite{farhi2014quantum} and adiabatic computation~\cite{farhi2000quantum}. Instead of using only two (non-commuting) operators as in QAOA, HVA uses more terms of the Hamiltonian. More specifically, 
\begin{equation}
    H = \sum_s H_s,\label{eq:Hdecomp}
\end{equation}
where we assume that each pair of $H_s$ and $H_{s'}$ do not commute, i.e., $[H_s, H_{s'}] \neq 0$. A depth $p$ HVA is given by
\begin{equation}\label{eq:hva}
    \ket{\psi_p} = \prod_{l=1}^p \left(\prod_s \exp{-i \theta_{s,l} H_s}\right)\ket{\psi_0},
\end{equation}
where $\ket{\psi_0}$ is the ground state of one of the terms in \cref{eq:Hdecomp}, i.e. $H_{s_0}$. When ordering the unitaries, we make sure that $H_{s_0}$ is not the first $H_s$ acting on $\ket{\psi_0}$. Note that due to the periodicity of the complex exponent, we can restrict the parameters to $[0, 2\pi]$, although in the case of certain symmetries, this restriction can be made tighter without losing expressive power~\cite{Ho2019}. Since these circuits are model-specific, the properties of the circuit can vary per problem. We give some concrete examples of HVA in the next section.

\section{Methods \& Models}\label{sec:methods}

    \subsection{Models}\label{sec:models}
        \subsubsection{Transverse Field Ising-Model}
        The TFIM is a paradigmatic model for studies of quantum magnetism. The Hamiltonian for the one-dimensional chain is given by:
        \begin{align}
            H_{\text{TFIM}} = -\sum_{i=1}^N\left[ \sigma_i^z \sigma_{i+1}^z + g \sigma_i^x \right]= H_{zz} + g H_x \label{eq:hamiltonian_TFIM},
        \end{align}
        with 
        $H_{zz} = -\sum_{i=1}^N \sigma_i^z \sigma_{i+1}^z$ and $H_x= - \sum_{i=1}^N \sigma_i^x$
        where we assume $g>0$ and use periodic boundary conditions $\sigma_{N+1}^{z}\equiv \sigma_1^{z}$. Here, $\sigma_{i}^\alpha$ corresponds to a Pauli matrix $\alpha=x,y,z$ acting on a site $i$, where the Pauli matrices are defined as follows:
    \begin{align*}
        \sigma_x &\equiv \begin{pmatrix}
                        0 & 1\\
                        1 & 0  
                \end{pmatrix} \quad
        \sigma_y \equiv \begin{pmatrix}
                        0 & -i\\
                        i & 0  
                \end{pmatrix} \quad
        \sigma_z \equiv \begin{pmatrix}
                        1 & 0\\
                        0 & -1  
                \end{pmatrix}.
    \end{align*} 
         The Hamiltonian has a $\mathbb{Z}_2$ symmetry so it is invariant under the operation of flipping all spins.
        
        For $g<1$, the system is in a ferromagnetic phase where the Hamiltonian favors spin alignment along the $z$ direction. For $g>1$ the system transitions to a disordered paramagnetic phase. In the limit that $g\to \infty$, the $\sigma^x$ term dominates the Hamiltonian, and the ground state becomes $\ket{+}^{\otimes N}$. At $g=1$ there is a critical point, and the system becomes gapless in the thermodynamic limit.
        
\begin{figure}[htb!]
            \centering
            \includegraphics[width=0.6\columnwidth]{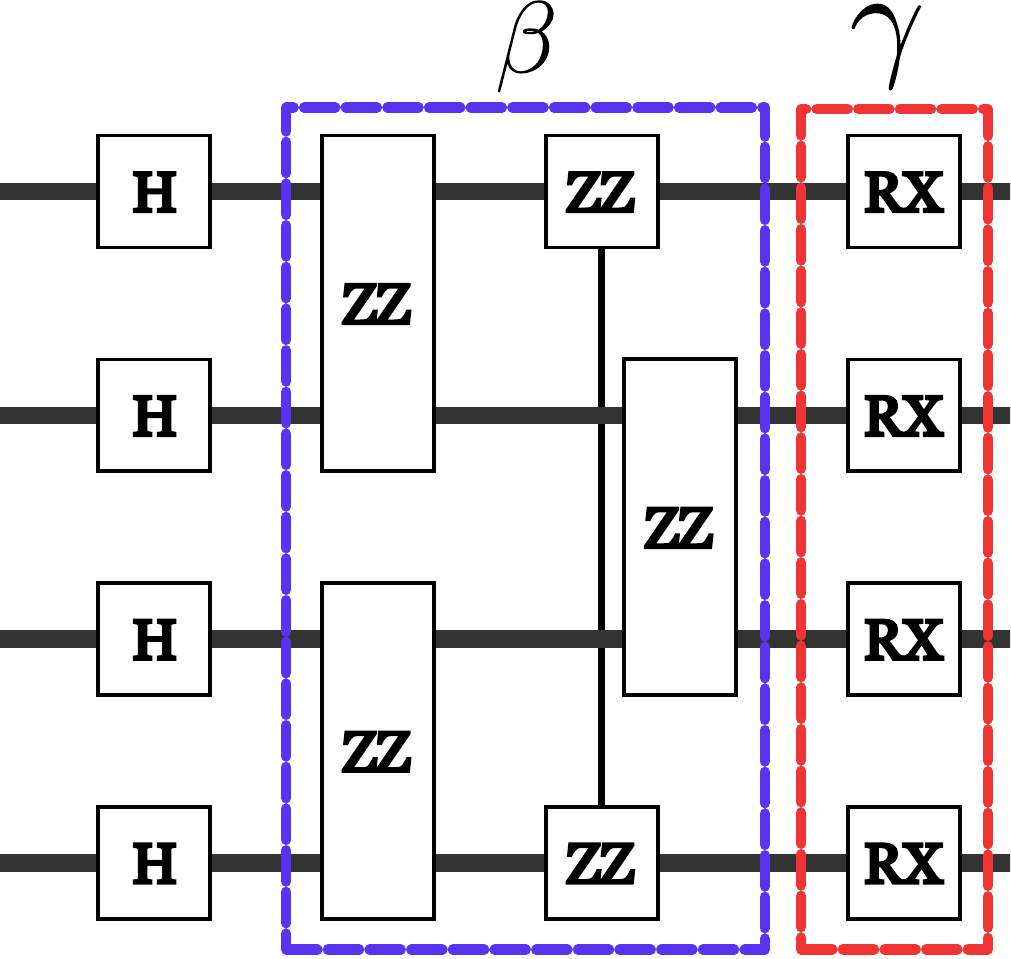}
            \caption{HVA quantum circuit for TFIM with $p=1$. The first layer of Hadamard gates, represented by H, are used to construct the initial $\ket{+}$ state. The ZZ gates are 2-local qubit rotation gates of the form $\text{ZZ} = \exp{ i\beta_l/2\: \sigma_i^z\sigma_j^z }$}. The RX gates are single qubit rotation gates $\text{RX}=\exp{i\gamma_l/2 \:\sigma_i^x}$.
            \label{fig:hva_tfi}
        \end{figure}

        A depth-$p$ HVA circuit for the TFIM corresponds to 
        \begin{align}
            U_{\text{TFIM}}(\bm{\beta},\bm{\gamma}) = \prod_{l=1}^p \exp \left\{ -i\frac{\beta_l}{2} H_{zz} \right\} \exp \left\{ -i\frac{\gamma_l}{2}  H_x \right\}. \label{eq:qaoa_tfi}
        \end{align}
        Hence for a depth $p$ circuit, we have $2p$ parameters. \Cref{fig:hva_tfi} illustrates the corresponding quantum circuit for $N=4$ and $p=1$. Note that we choose $\ket{\psi_0}$ in \eqref{eq:hva} to be the ground state of $H_x= - \sum_{i=1}^N \sigma_i^x$, i.e., $\ket{\psi_0} = \ket{+}^{\otimes N}$. The HVA circuit of \eqref{eq:qaoa_tfi} is the same as the QAOA ansatz used in~\cite{farhi2014quantum} for solving the MaxCut problem. By using the Jordan-Wigner transformation, it was shown that the ground state can be represented accurately with a depth $p = N/2$ circuit for the case that $g=0$~\cite{Mbeng2019}. For the case that $g\neq0$, there is only numerical evidence to support this claim~\cite{Ho2019, wierichs2020avoiding}. In appendix \cref{app:numres} we confirm that for the TFIM one can consistently find the ground state for $g\in\{0.5, 0.52,\ldots,1.5\}$ with a depth $p=N/2$ circuit. 
        
        \subsubsection{XXZ-model}
        Another prototypical model for studying quantum magnetism is the XXZ model.
        For the 1D XXZ model, the Hamiltonian is given by
        \begin{align}\label{eq:ham_xxz}
            H_{\text{XXZ}} &= \sum_{i=1}^N\left[\sigma_i^x \sigma_{i+1}^x +  \sigma_i^y \sigma_{i+1}^y  +  \Delta \sigma_i^z \sigma_{i+1}^z \right] \nonumber \\ 
            &= H_{xx} + H_{yy} + \Delta H_{zz}
        \end{align}
        with $H_{xx} = \sum^N_{i=1}\sigma_i^x \sigma_{i+1}^x$, $H_{yy} = \sum^N_{i=1}\sigma_i^y \sigma_{i+1}^y$ and $H_{zz} = \sum^N_{i=1}\sigma_i^z \sigma_{i+1}^z$. Again we assume periodic boundary conditions. The parameter $\Delta$ controls the spin anisotropy in the model. For $\Delta=1$, this model has an $SU(2)$ symmetry and is equivalent to the Heisenberg chain. For $\Delta\neq1$, this symmetry gets reduced to a $U(1) \times \mathbb{Z}_2$ symmetry. 
        For $1<\abs{\Delta}$ the system is in the XY quasi-long-range ordered state and becomes gapless in the  thermodynamic limit. 
        At $\abs{\Delta}=1$ there is a phase transition to the N\'eel ordered state. This model can be solved exactly using the Bethe-ansatz for $N\to \infty$~\cite{Franchini2017integrable}. 
        
\begin{figure}[htb!]
            \centering
            \includegraphics[width=0.95\columnwidth]{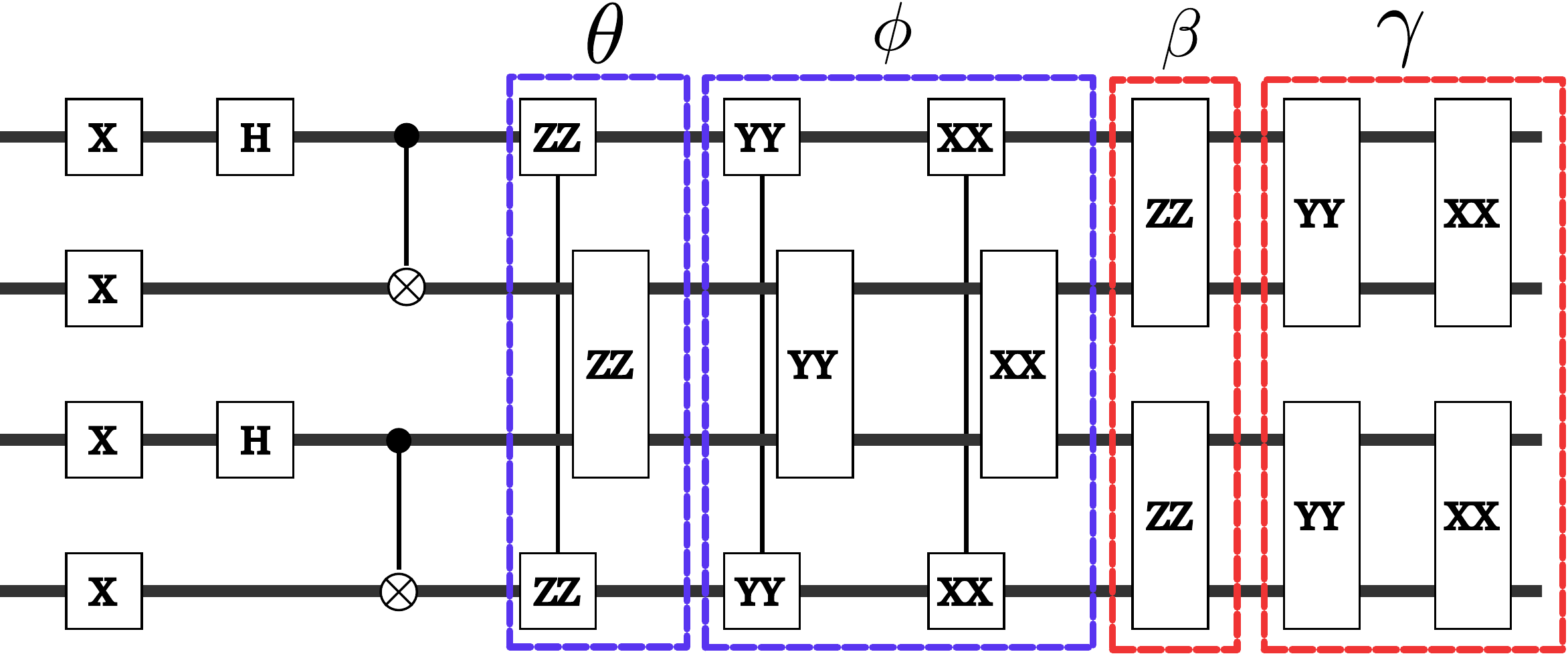}
            \caption{HVA quantum circuit for the XXZ model with $p=1$. Here, the X gates are given by $\text{X} = \sigma_i^x$. Together with a single Hadamard gate and a CNOT on even links, we prepare the $\ket{\Psi^-}$ Bell state. The 2-local qubit rotations are all of the form $\text{AA} = \exp{ -i x/2\: \sigma_i^a\sigma_j^a }$, with $x=\theta,\phi ,\beta, \gamma$ depending on whether the links are even or odd and ZZ or XX, YY.}
            \label{fig:hva_xxz}
        \end{figure}

        Inspired by~\cite{Ho2019}, we decompose the 1D chain into even and odd links and separate the Hamiltonian into two parts, 
        \begin{align*}
            H^{\text{even}} &= H^{\text{even}}_{xx} + H^{\text{even}}_{yy} + H^{\text{even}}_{zz} \\ 
            H^{\text{odd}}  &= H^{\text{odd}}_{xx} + H^{\text{odd}}_{yy} + H^{\text{odd}}_{zz} ,
        \end{align*}
        where the indices only run over non-overlapping bonds:
        \begin{align*}
            H^{\text{even}}_{\alpha\alpha} = \sum_{i=1}^{N/2} \sigma_{2i-1}^\alpha \sigma_{2i}^\alpha \quad \text{ and } \quad
            H^{\text{odd}}_{\alpha\alpha} = \sum_{i=1}^{N/2} \sigma_{2i}^\alpha \sigma_{2i+1}^\alpha
        \end{align*}
        for $\alpha =x,y,z$. Our numerical experiments indicate that separately parameterizing these bonds gives better performance when studying the anisotropic system $\Delta\neq1$. Additionally, we parameterize $H_{xx},H_{yy}$ and $H_{zz}$ terms with their own respective parameter. 
        The reason for this is that for $\Delta\neq 1$ the anisotropy in the model cannot be accounted for by a single parameter.
        A depth-$p$ HVA circuit for the XXZ model corresponds to 
        \begin{align}\label{eq:qaoa_xxz}
            U_{\text{XXZ}}(\bm{\beta},\bm{\gamma}) & = \prod_{l=1}^p  G(\theta_l, H_{zz}^{\text{odd}}) G(\phi_l, H_{xx}^{\text{odd}}) G(\phi_l, H_{yy}^{\text{odd}})\nonumber\\
            & \quad G(\beta_l, H_{zz}^{\text{even}}) G(\gamma_l,H_{xx}^{\text{even}}) G(\gamma_l, H_{yy}^{\text{even}})
        \end{align}
        where $$G(x, H) =\exp{-i \frac{x}{2} H}.$$
        Hence for a depth-$p$ circuit, we have $4p$ parameters. \Cref{fig:hva_xxz} illustrates a quantum circuit for $N=4$ and $p=1$. We choose the initial state $\ket{\psi_0}$ in \eqref{eq:hva} to be the ground state of $H^{\text{even}}$, i.e., $\ket{\psi_0} = \bigotimes^{N/2}_{i=1} \frac{1}{\sqrt{2}} \left(\ket{01} - \ket{10}\right)_{2i-1, 2i} =\bigotimes^{N/2}_{i=1}\ket{\Psi^-}$. It was shown in~\cite{Ho2019} that the Heisenberg chain (i.e., $\Delta = 1$) can be solved accurately with HVA with $p = N/2$. Note that for the case of $\Delta = 1$, one can use a single parameter for $H_{xx}+H_{yy}+H_{zz}$. In appendix \cref{app:numres}, we find that for $\Delta \in \{0.5, 0.52, \ldots, 1.5\}$ and $\Delta \neq 1$, a depth $p = N/2$ HVA circuit is sufficient to find a close approximation to the ground state.
        
    In this work, we consider the problem of approximating the ground state at the critical points $g=1$ and $\Delta=1$ for the TFIM and XXZ model respectively since their particular entanglement scaling properties makes them harder to approximate with classical methods~\cite{orus2019}, such as density matrix renormalization group (DMRG). Due to the criticality of the aforementioned systems at these order values, the energy spectrum becomes gapless in the thermodynamic limit and hence there is a logarithmic correction of $S \propto \log{N}$ to the area law of entanglement entropy. A matrix product state with bond dimension $D$ bounds the entanglement of the state to $S\leq 2\log{D}$, so the necessary bond dimension to express the ground state grows polynomially in a DMRG calculation~\cite{orus2019}.

    \subsubsection{Performance Metrics}
    We use the \emph{fidelity} $\mathcal{F}$ between the VQE optimized state $\ket{\psi(\bm{\theta}^*)}$ and the true ground state $\ket{\psi_{\text{ground}}}$ obtained from exact diagonalization:
    \begin{align*}
        \mathcal{F} = \abs{\braket{\psi(\bm{\theta^*})}{\psi_{\text{ground}}}}.
    \end{align*}
    Note that for the models studied in this work, $\ket{\psi_{\text{ground}}}$ is always non-degenerate. If the fidelity is $>99.9\%$, we assume that we have successfully found the ground state. When assessing the quality of an optimized HVA circuit, the fidelity is a strong indicator of the success for solving the ground state problem, since the infidelity upper bounds the difference between the ground state and variational expectation value of any observable. Let $1-\mathcal{F} < \epsilon$
    \begin{align*}
        \abs{\expval{O}_{\text{ground}}-\expval{O}_{\bm{\theta}}} \leq 2c\sqrt{\epsilon(1-\epsilon)} + \epsilon
    \end{align*}
    where $c$ is the operator norm of $O$~\cite{Beach2019sprint}.

\subsection{Entanglement}
In the context of quantum many-body physics, quantum correlations play a central role in our current understanding  of the equilibrium and out-of-equilibrium properties of several systems in condensed matter. The source of these correlations is inherently non-local, and can be traced back to the presence of entanglement in the quantum state. In this section we introduce several commonly used entanglement quantities in quantum many-body physics.

\begin{figure}[htb!]
    \centering
    \includegraphics[width=0.9\columnwidth]{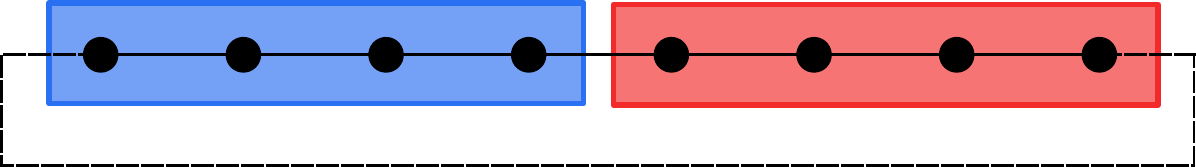}
    \caption{Division of the full system into two subsystems $A$ (blue) and $B$ (red) on a one-dimensional chain.}
    \label{fig:chain_cut_TFI}
\end{figure}

In classical systems, one uses entropy to quantify our lack of knowledge of the state of the system due to thermal fluctuations. However, for a quantum system at zero temperature, the entropy of a subsystem has a different origin: entanglement. To quantify it, we use the bipartite \emph{entanglement entropy}~\cite{Eisert2010arealaw}, which is defined as the \emph{von Neumann entropy} of the reduced density matrix $\rho_A$. To obtain this reduced density matrix, we divide the system into two subsystems $A$ and $B$ and trace out subsystem $B$,
\begin{equation}
    \rho_A (\ket{\psi}) = \Tr_B(\ket{\psi}\bra{\psi}).
\end{equation}
where $\ket{\psi}$ is a pure state. For example, for an 8-spin model on a ring, a typical bipartition is given in \cref{fig:chain_cut_TFI}. 

\noindent The von Neumann entropy generalizes the concept of Shannon entropy to quantum states, and is given by
\begin{equation}
    S(\rho_A) \triangleq -\Tr(\rho_A \log{\rho_A})
\end{equation}
Since a bipartite quantum state can always be rewritten using the Schmidt decomposition,
\begin{align}
    \ket{\psi} = \sum_{k=0}^K e^{-\frac{1}{2}\xi_k} \ket{\psi^k_A}\otimes\ket{\psi^k_B},\label{eq:schmidt}
\end{align}
with $\braket{\psi^k_A}{\psi^m_A}=\braket{\psi^k_B}{\psi^m_B}=\delta_{km}$ and $K$ the size of the smallest subsystem, the von Neumann entropy reduces to~\cite{Li2008entspec}
\begin{equation}
    S(\rho_A)  = \sum_{k=0}^K \xi_k \exp{-\xi_k}\label{eq:vn_entropy},
\end{equation}
In recent years, the importance of entanglement in condensed matter physics has been elucidated in several systems through the study of the scaling behaviour of the entanglement entropy, which has enabled the identification and characterization of exotic phases of matter such as topological quantum states~\cite{Kitaev2006topological} and quantum spin liquids~\cite{Isakov2011spinliquid, Zhang2011critspinliq}.

Fully characterizing the entanglement properties of a system cannot be done by looking solely at the entanglement entropy~\cite{Li2008entspec, Yang2015mpdist, Shaffer2014irrcirc}. The so-called \emph{entanglement spectrum} has a much richer structure, and has been used to study many-body localization~\cite{Yang2015mpdist}, observable thermalization~\cite{Geraedts2016thermalize}, irreversibility in quantum circuits~\cite{Shaffer2014irrcirc}, and preparation of
ground states of non-integrable quantum models \cite{matos2020quantifying}. In addition, the entanglement spectrum has been used to study the properties of variational methods such as the Restricted Boltzmann Machine~\cite{Deng2017}. The entanglement spectrum is defined as the eigenvalue spectrum of the \emph{entanglement Hamiltonian}
\begin{equation}
    H_{\text{ent}} \triangleq - \log \rho_{A}. \label{eq:ent_spect}
\end{equation}
From \cref{eq:schmidt} it follows directly that this Hamiltonian has eigenvalues $\xi_k$. For random quantum states distributed according to the Haar measure, the entanglement spectrum follows the \emph{Marchenko-Pastur distribution}~\cite{Znidaric2006randomvec, marenko_mp_1967}. This distribution describes the asymptotic average density of eigenvalues of Wishart matrices, i.e., matrices of the form $XX^*$ where $X$ be $m \times n$ random matrices. 

Finally, the \emph{Page entropy}~\cite{page1993} describes the average entanglement entropy over randomly drawn pure states in the entire Hilbert space, and is given by
\begin{equation}
    S_{Page} = - \frac{d_A - 1}{2d_B} + \sum_{k=d_B + 1}^{d_Ad_B} \frac{1}{k}\approx \log(d_A) - \frac{d_A}{2d_B},
\end{equation}
where $d_A$ and $d_B$ are the dimensions of subsystem $A$ and $B$, respectively.

\section{Main Results}\label{sec:main}

\subsection{The ansatz space through the lens of entanglement spectrum}
\label{sec:ansatz-space}

The effectiveness of a VQE optimization is determined by two factors. First, one requires an expressive enough \emph{ansatz space} that contains the ground state. The ansatz space of a specific model $H$ and depth $p$ refers to the set of all possible quantum states that can be reached by applying a depth-$p$ HVA circuit corresponding to $H$ to a fixed initial state $\ket{\psi_0}$ which depends on the model. Secondly, the non-convex cost landscape induced by the variational energy of \cref{eq:cost} must be favorable, in the sense that the optimization does not get stuck in local minima and can reliably reach the ground state. 

Here, we investigate the properties of the ansatz space by examining the entanglement spectra of HVA quantum states generated with random parameters sampled uniformly in the range $[0,\pi]$ for the TFIM and $[0,2\pi]$ for the XXZ model. For each model, we sample $5000$ sets of parameters and calculate the entanglement spectrum of the resulting state. If the spectrum of the sampled states follows a distribution close to the Marchenko-Pastur (MP) distribution, a random HVA state has entanglement spectrum that resembles that of a Haar random state. On the contrary, a distribution far away from the MP distribution indicates a restricted manifold of states that has a non-random structure. We hypothesize that the shape of the average entanglement spectrum can give insight into the performance of the VQE optimization by revealing the structure of the ansatz space.

\begin{figure}[htb!]
\subfloat[\label{fig:tfimp16} TFIM]{%
    \centering
    \includegraphics[width=0.9\columnwidth]{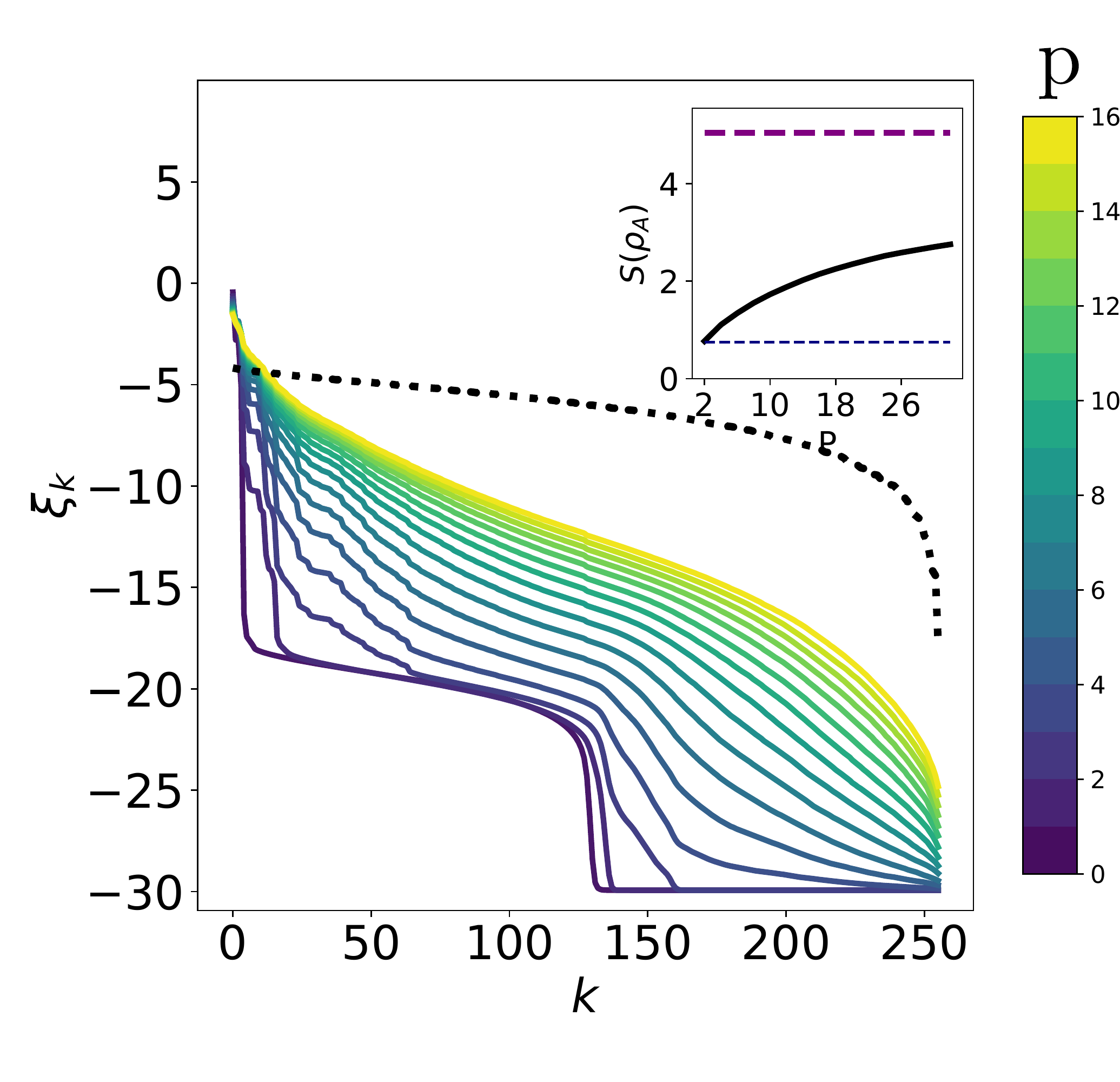}
}

\subfloat[\label{fig:xxzmp16} XXZ model]{
    \centering
    \includegraphics[width=0.9\columnwidth]{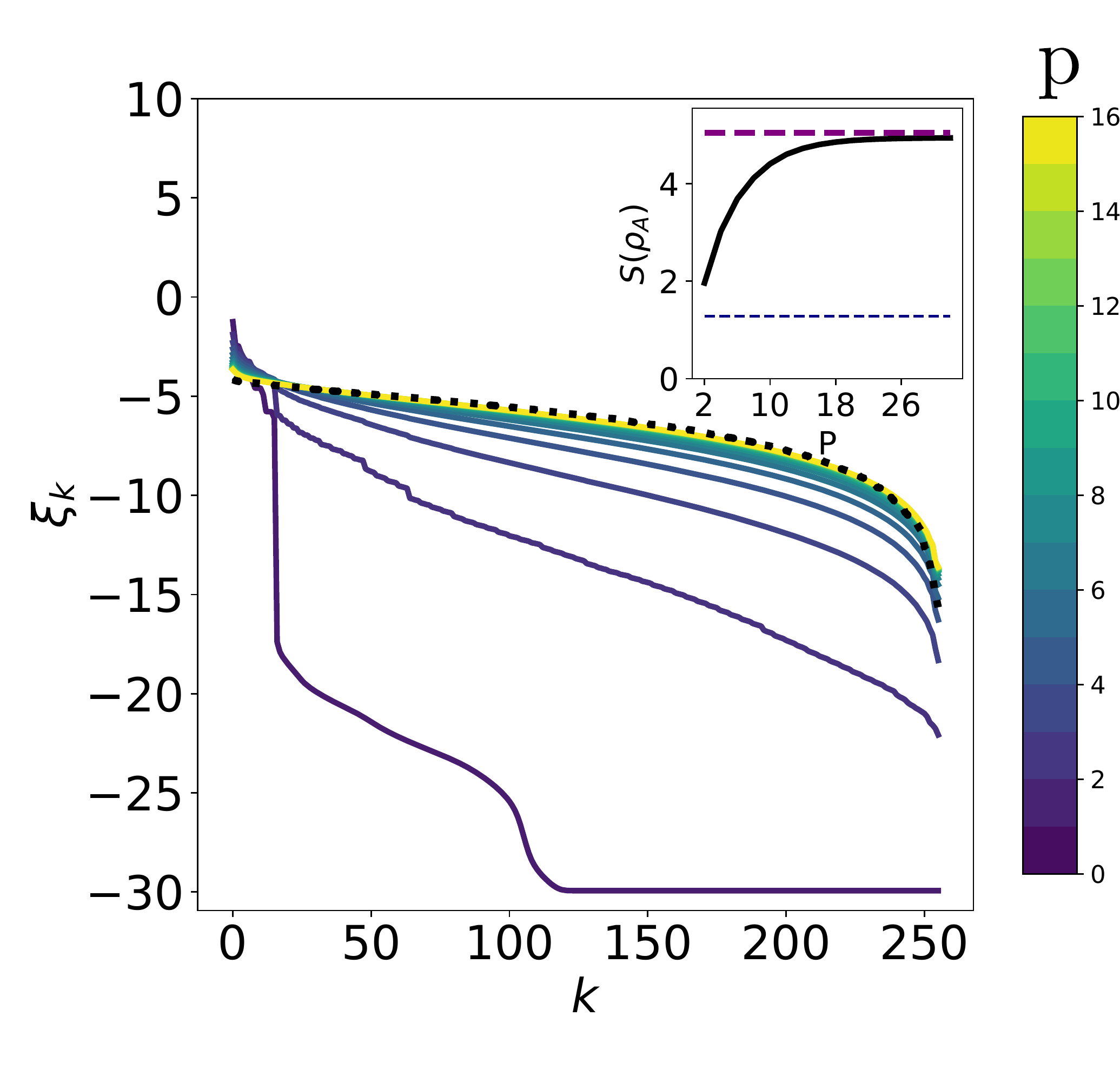}
}
\caption{Average entanglement spectrum of HVA quantum states from layer $p=1$ (bottom line in purple) to $p=N$ (top line in yellow) over $5000$ random parameter initializations. $\xi_k$ denotes the $k$-th eigenvalue of $H_{\text{ent}}$. The eigenvalues are arranged in descending order and cut off at $\xi_k=-30$. The black lines in the insets show how close the average entanglement entropy is to the Page-entropy (purple dashed line) as a function of increasing circuit depth. The lower blue dashed line in the inset indicates the entanglement entropy of the ground state. We see that the average HVA state is more entangled than the ground states of interest.}
\label{fig:mp}
\end{figure}

\Cref{fig:mp} shows the average entanglement spectrum for a state in the ansatz space of circuits with depths ranging from $1,2,\ldots,N$ for the $N=16$-qubit TFIM and XXZ models. From the insets we see that both ans\"atze have enough entangling power to express the ground state, even for low depth circuits. For the TFIM with 16 qubits (\cref{fig:tfimp16}), we see that for all $p$ the HVA spectrum is further away from the MP distribution, and the HVA space corresponding to the TFIM appears to be a manifold of states with restricted entanglement structure. In contrast, for the XXZ model, we see that the average spectra gets closer to the MP distribution as $p$ increases. This suggests that the HVA space for the XXZ model is not as restricted as for the TFIM. This can be understood directly by looking at the circuit complexity, which for the XXZ model contains more gates and parameters per layer. However, this is necessary because the XXZ model is inherently a much richer model in terms of physics, and it may be necessary for HVA space to accommodate more variety of states.

We now turn to examining the entanglement features of the XXZ model HVA states explored during optimization. For the variational minimization of \cref{eq:cost} we use a gradient descent algorithm (see \cref{app:compdet} for details). Since the cost function is non-convex, the quality of the solution will vary significantly between different starting points in parameter space. We compare the following initialization strategies:
\begin{enumerate}
    \item A completely random-state initialization, where all parameters are sampled as $\bm{\theta}\sim\mathcal{U}(0,2\pi)$.
    \item An identity initialization. We set all parameters equal to $\pi$, so that our circuit is equal to the identity circuit and a global phase $i$. 
\end{enumerate}
Our approach of starting close to the identity is similar to the block identity initialization strategy discussed in~\cite{Grant2019initialization}, however, we study a simpler version by setting all parameters equal to $\pi$. For both parameter initializations, we extract the final layer state from the circuit at multiple times during the optimization and calculate its entanglement spectrum with \cref{eq:ent_spect}. Not surprisingly, our experiments indicate that a random start is prone to getting stuck in a local minimum, due to our local optimization strategies combined with a non-convex energy landscape. The identity start on the other hand allows one to consistently find a high fidelity state for both systems with a depth $p=N/2$ circuit (see appendix \cref{app:numres}). 

\begin{figure}[htb!]
\subfloat[\label{fig:xxzmp16regularstart}]{
    \centering
    \includegraphics[width=0.9\columnwidth]{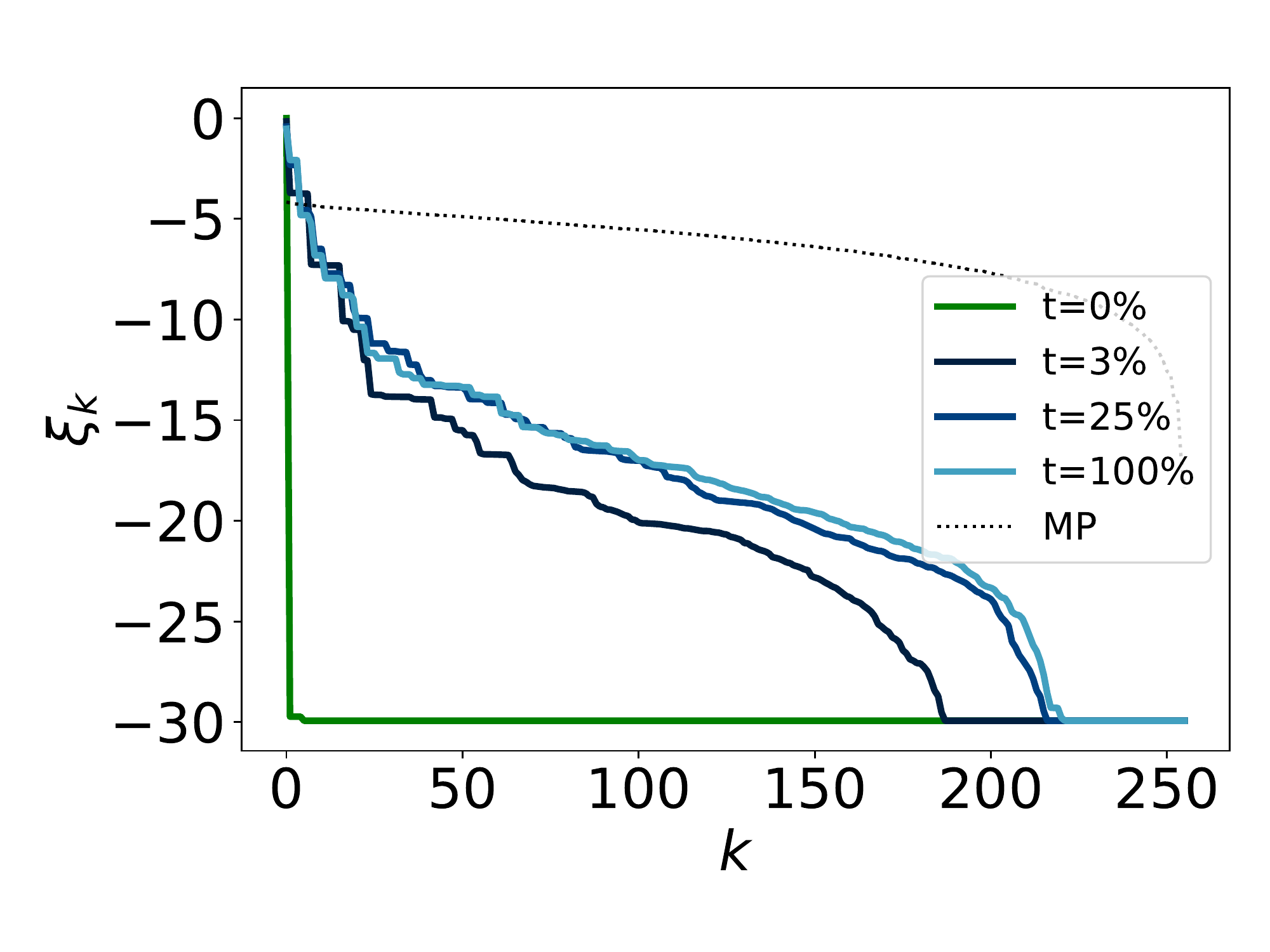}
}

\subfloat[\label{fig:xxzmp16randomstart}]{
    \centering
    \includegraphics[width=0.9\columnwidth]{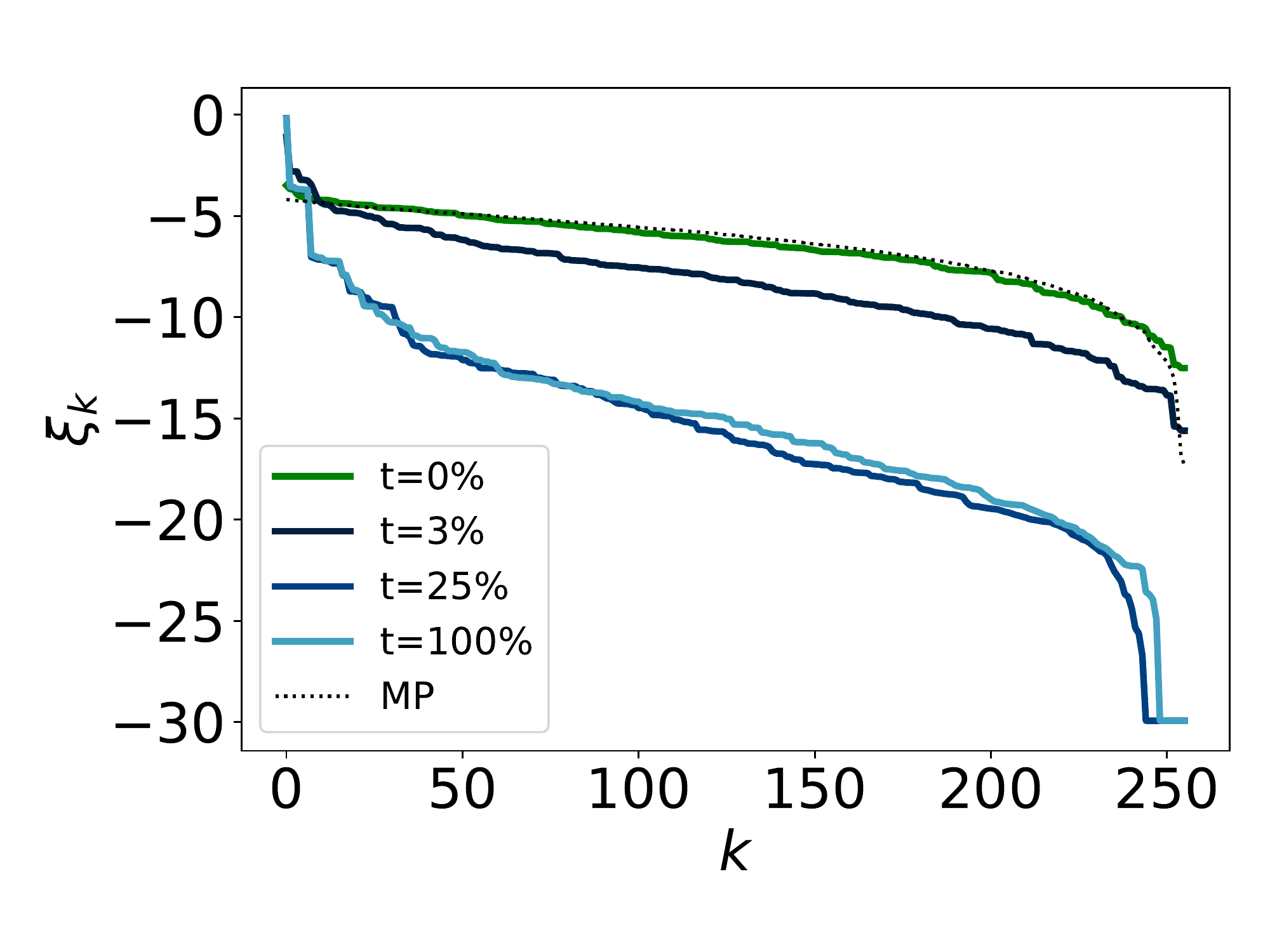}
}
    \caption{Change of the entanglement spectrum of the final layer during the optimization. Both figures are for a 16 qubit XXZ model with a depth $p=N/2$ circuit. The times are percentages of the total optimization time. Figure (a) correspond to a converged state of  fidelity, whereas figure (b) corresponds to a $\approx70\%$ fidelity state. (a) The identity state initialization remains far away from the MP distribution at all times during the optimization and convergence to state with $>99.9\%$ with the ground state. Since this initialization strategy starts with the identity circuit, we find the $t=0\%$ state to be a product state, as indicated by the single eigenvalue. (b) The random initialization starts close to the MP distribution and converges to a local minimum with $\approx70\%$ fidelity.}
    \label{fig:mp_dist_dynamics}
\end{figure}

To study this finding in more detail, we study the dynamics of the entanglement spectrum for different initialization strategies. In \cref{fig:xxzmp16regularstart} we see that an identity state initialization stays far away from the MP distribution at all times, indicating that we are accessing a highly structured restricted subspace of the full HVA space. Additionally, this initialization reaches a state with $>99.9\%$ fidelity state. On the contrary, the random state initialization in \cref{fig:xxzmp16randomstart} starts close to the MP distribution and then moves to a more structured, local minima with $70\%$ fidelity. We conclude that even though the shape of the entanglement spectrum from \cref{fig:xxzmp16} indicates a possible large unstructured ansatz space, a local optimization is still capable of finding the ground state if we choose a suitable parameter initialization. We further investigate the qualitative properties of the optimization dynamics in appendix \cref{app:entropydynamics}. In the next section, we will see that the disadvantage of starting at a bad initial point can be overcome by making the circuit sufficiently deep, a process known as \emph{over-parameterization}. 

\subsection{Over-parameterization in HVA}\label{sec:overparam}

Over-parameterization is a phenomenon in certain types of non-convex optimization problems. For an over-parameterized model, the optimization landscape becomes dramatically better (e.g., almost trap free or almost-convex) as the number of parameters reaches some threshold. In most cases the rate of convergence also becomes better, sometimes even exponentially faster after passing this threshold. 

\begin{figure}[htb!]
\centering
\subfloat[\label{fig:tfioverparam} TFIM]{
    \centering
    \includegraphics[width=0.9\columnwidth]{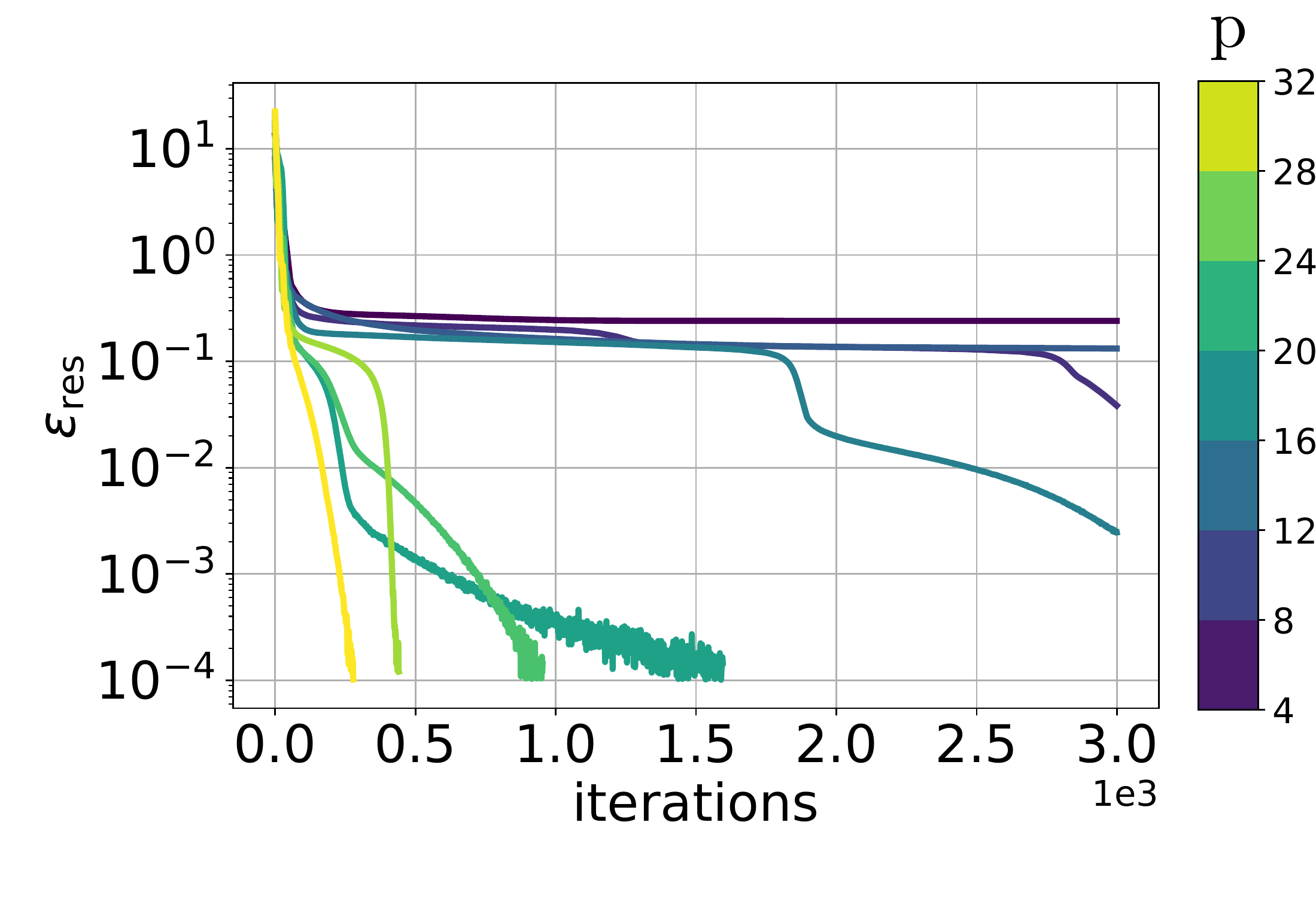}
}

\subfloat[\label{fig:xxzoverparam} XXZ model]{
    \centering
\includegraphics[width=0.9\columnwidth]{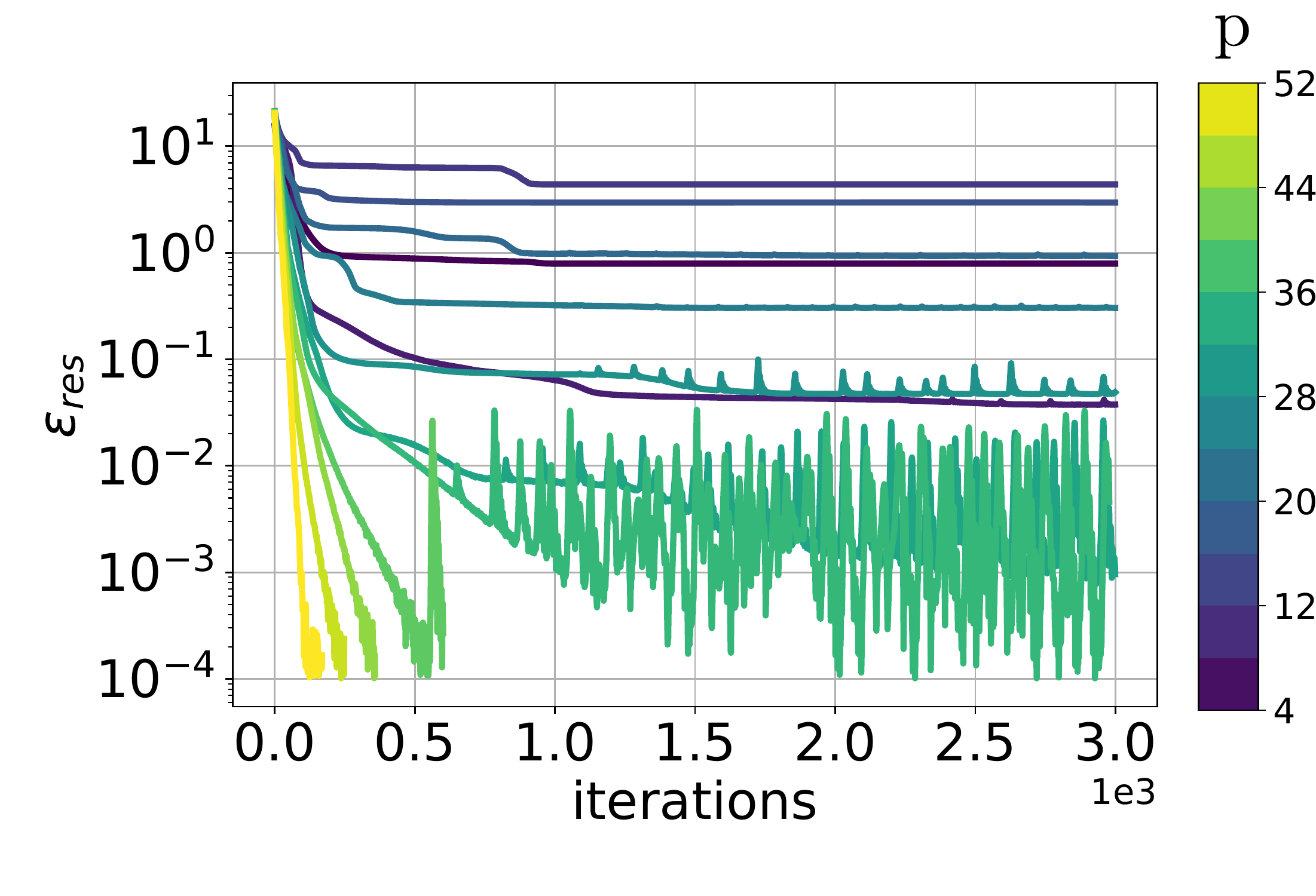}
}
\caption{Over-parameterization in HVA. Each line corresponds the VQE optimization at depth $p$ that took the most iterations to converge out of 100 random initializations. Both figures are for $N=12$ qubits. The rapid oscillations in figure (b) are artifacts of the Adam optimizer and are less severe as the circuit depth increases. Due to our stopping criterion, we know that if the number of iterations is smaller than $3000$, then $\epsilon_{\text{res}}\leq1e-4$ and so the model did converge to a good ground state approximation.}
\label{fig:overparam}
\end{figure}

Over-parameterization has been studied extensively in the classical deep neural network literature~\cite{allenzhu2018learning, allenZhu2019over, chen2019overparameterization, chen2020meanfield}. For example, in~\cite{allenZhu2019over} it was shown that under certain mild assumptions, the optimization landscape of a deep neural network is almost-convex in a large neighborhood of a random starting point. As a consequence the stochastic gradient descent algorithm can almost always find an accurate solution. 

Although for VQE algorithms it is clear that we have $\min E(\bm{\theta})_{p+1}\leq \min E(\bm{\theta})_{p}$, it is not clear if this minimum can be found consistently due to the non-convexity of the energy landscape. Hence, a deeper understanding of the energy landscape with increasing depth is required. There is some work on over-parameterization in the context of controllable quantum systems with \emph{unconstrained} time-varying controls~\cite{rabitz1998optimalcontrol,rabitz05optimalcontrol, russell2017control}, where the authors show that there are no sub-optimal local minima in the optimization landscape. For the case of a \emph{constrained} controllable quantum system, a recent work~\cite{kiani2020learning} considers the problem of learning $d$-dimensional Haar random unitaries $U(d)$ by gradient descent using general alternating operator ansatz of the form $e^{-i\gamma_p A}e^{-i\beta_p B}\cdots e^{-i\gamma_1 A}e^{-i\beta_1 B}$, where $A$ and $B$ are matrices sampled from the Gaussian Unitary Ensemble~\cite{LalMehta2004irmt}. The authors show that gradient descent always converges to an accurate solution when the number of parameters is $d^2$ or greater, and a ``computational phase transition'' is observed between an under-parameterization ($<d^2$) and over-parameterization ($>d^2$) regimes.

Since HVA also has the form of an alternating operator ansatz, and the problem of finding the ground states can also be seen as a constrained quantum control problem, we expect a similar over-parameterization phenomenon in our setting. To investigate this, we randomly sample 100 initial parameters $\bm{\theta}$ (uniformly drawn from the interval $[0,\pi]$ for the TFIM and $[0,2\pi]$ for the XXZ model) and perform the optimization for increasing values of $p$. Here, we set the stopping criterion for the optimization to $\epsilon_{\text{res}} =E(\bm{\theta})_p- E_{\text{ground}}<1e-4$ and the maximum number of iterations to 3000. Indeed, \cref{fig:overparam} shows that the over-parameterization phenomenon also occurs in HVA for the 12-qubit TFIM and XXZ model. We find that for both the TFIM and XXZ model, gradient descent from all 100 random starting points converges to an accurate solution once the depth $p$ reaches a certain threshold $\tilde{p}(N)$.

\begin{figure}[htb!]
\subfloat[\label{fig:scalingtfioverparam} TFIM]{
    \centering
    \includegraphics[width=0.9\columnwidth]{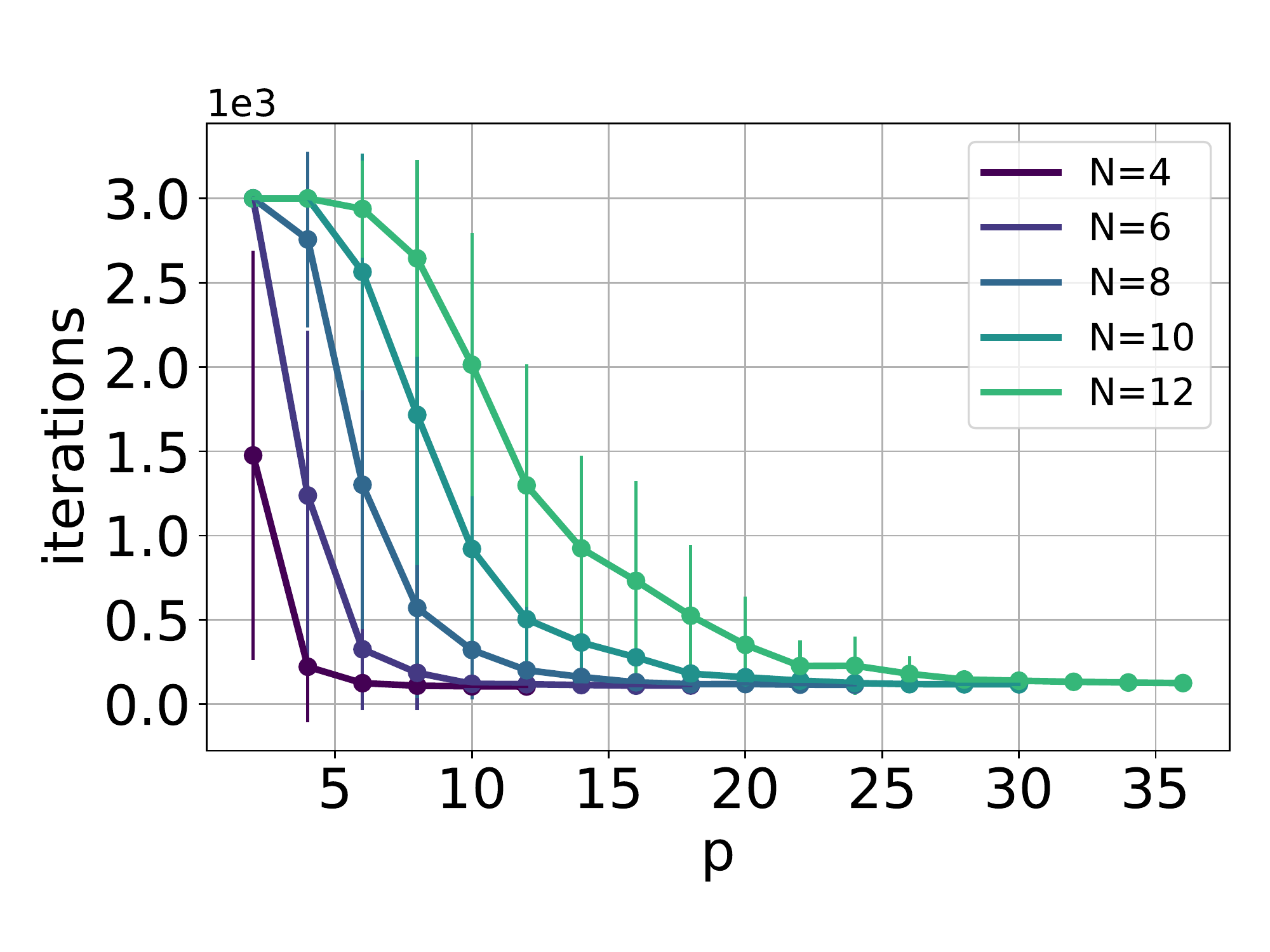}
}

\subfloat[\label{fig:scalingxxzoverparam} XXZ model]{
    \centering
    \includegraphics[width=0.9\columnwidth]{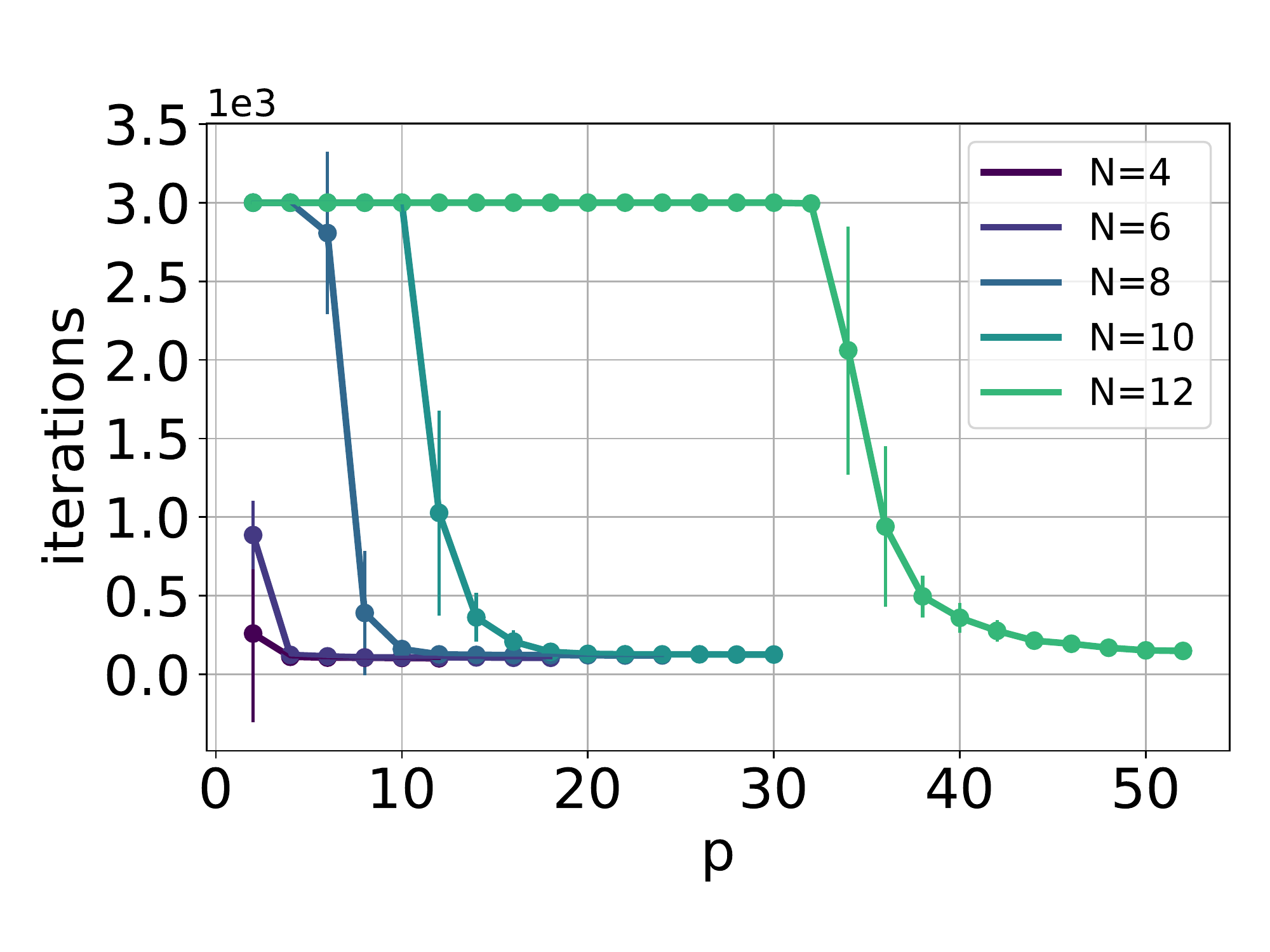}
}
\caption{The mean iteration time to convergence as a function of depth. Error bars indicate the standard deviation over 100 different initializations. For both models there is a clear cutoff where the number of iterations saturate. Note that if the number of iterations is smaller than $3000$, then we know that $\epsilon_{\text{res}} \leq 1e-4$, indicating that the optimization has converged to a good ground state approximation. We see that the error bars decrease systematically with depth. For both models, there is a critical $p$ after which all random initializations converge to a good ground state approximation. Moreover, for depth $p=34$ and $p=52$ for the TFIM and XXZ model respectively, the number of iterations to find the ground state is of the order of 100 iterations for every starting point.}
\label{fig:scalingoverparam}
\end{figure}

Moreover, we also observe a ``computational phase transition'' around this threshold where the convergence speed becomes exponentially fast, i.e., the decrease of the residue energy as a function of the number of iterations. However, this threshold $\tilde{p}(N)$ is not tight, i.e., for depth $p<\tilde{p}(N)$ it is possible that gradient descent still converges to a high fidelity state. This indicates that in the setting of finding ground states using HVA, the problem is more structured and gradient descent is effective. In \cref{fig:scalingoverparam} we see that for all system sizes, the mean number of iterations eventually converges to about 100 iterations. In addition, we can find the over-parameterization thresholds $\tilde{p}(N)$ in \cref{tab:optable} for the TFIM and XXZ model with different system sizes. Our data suggests that $\tilde{p}(N)$ has at most a polynomial scaling, which is compatible with the analogous parameter count required to express critical 1D ground states with an MPS. A more detailed view of this data can be found in \cref{fig:percentage} in the Appendix, which shows that all random initializations converge to the ground state after a certain threshold $\tilde{p}(N)$.

This is a striking difference with~\cite{kiani2020learning} where the number of parameters to achieve over-parameterization is $(2^{N})^2$. From \cref{fig:scalingoverparam} we can also see that the iteration time decreases substantially as $p$ increases, which saturates to around $100$ iterations after a certain $p$ for all $N$.

The over-parameterization phenomenon in HVA shows a clear difference between HVA and parameterized random quantum circuits (RQC), because there is no indication or evidence that the landscape of RQC gets better as one increases the depth. On the contrary, in our experiments with random circuits of comparable depths to our HVA circuits, we were unable to observe the same  over-parameterization phenomenon. This can be explained from the barren plateau point of view and the lack of structure in the ansatz space.

\begin{table}[htb!]
    \centering
    \begin{tabular}{c|c|c}
            & TFIM & XXZ model\\\hline
        $N$ & $\tilde{p}(N)$ & $\tilde{p}(N)$ \\\hline 
        4 & 6 & 4\\ \hline 
        6 & 6 & 4\\ \hline
        8 & 8 & 8\\ \hline 
        10 & 10 & 12\\ \hline
        12 & 14 & 36\\ \hline
    \end{tabular}
    \caption{Over-parameterization threshold $\tilde{p}(N)$ for TFIM and XXZ model with different system sizes $N$. By threshold we mean that when $p \geq \tilde{p}(N)$, all the random initializations converged to an accurate solution.}
    \label{tab:optable}
\end{table}


\subsection{Ameliorated barren plateaus in HVA}\label{sec:nobarren}

In Ref.~\cite{McClean2018barren}, a \emph{barren plateau phenomenon} was observed for VQE on random quantum circuits, where all gradients are exponentially close to zero with overwhelmingly high probability, making local optimization within the ansatz space extremely challenging. The barren plateau phenomenon is due to the fact that RQCs consisting of single- and two-qubit gates form a $2$-design, which means that gradients of the energy objective function will obey the same concentration of measure properties as if the circuits were Haar-random unitaries.

In contrast to the RQC ansatz, we show that the optimization landscape of HVA is much more favorable. This is clearly illustrated when optimizing the HVA corresponding to the TFIM: to begin with, as discussed in \cref{sec:ansatz-space} the manifold of states has a much more restricted entanglement structure than a typical, Haar-random state -- this already indicates that the HVA circuits do not form $2$-designs, and thus do not obey the same kind of concentration of measure phenomenon as RQCs. On the other hand, the entanglement spectrum of the ansatz space corresponding to the XXZ model does not immediately rule out the same barren plateau behaviour as exhibited by RQCs.

\begin{figure}[htb!]
\centering
\subfloat[\label{fig:tfigrads16} TFIM]{
    \centering
    \includegraphics[width=0.95\columnwidth]{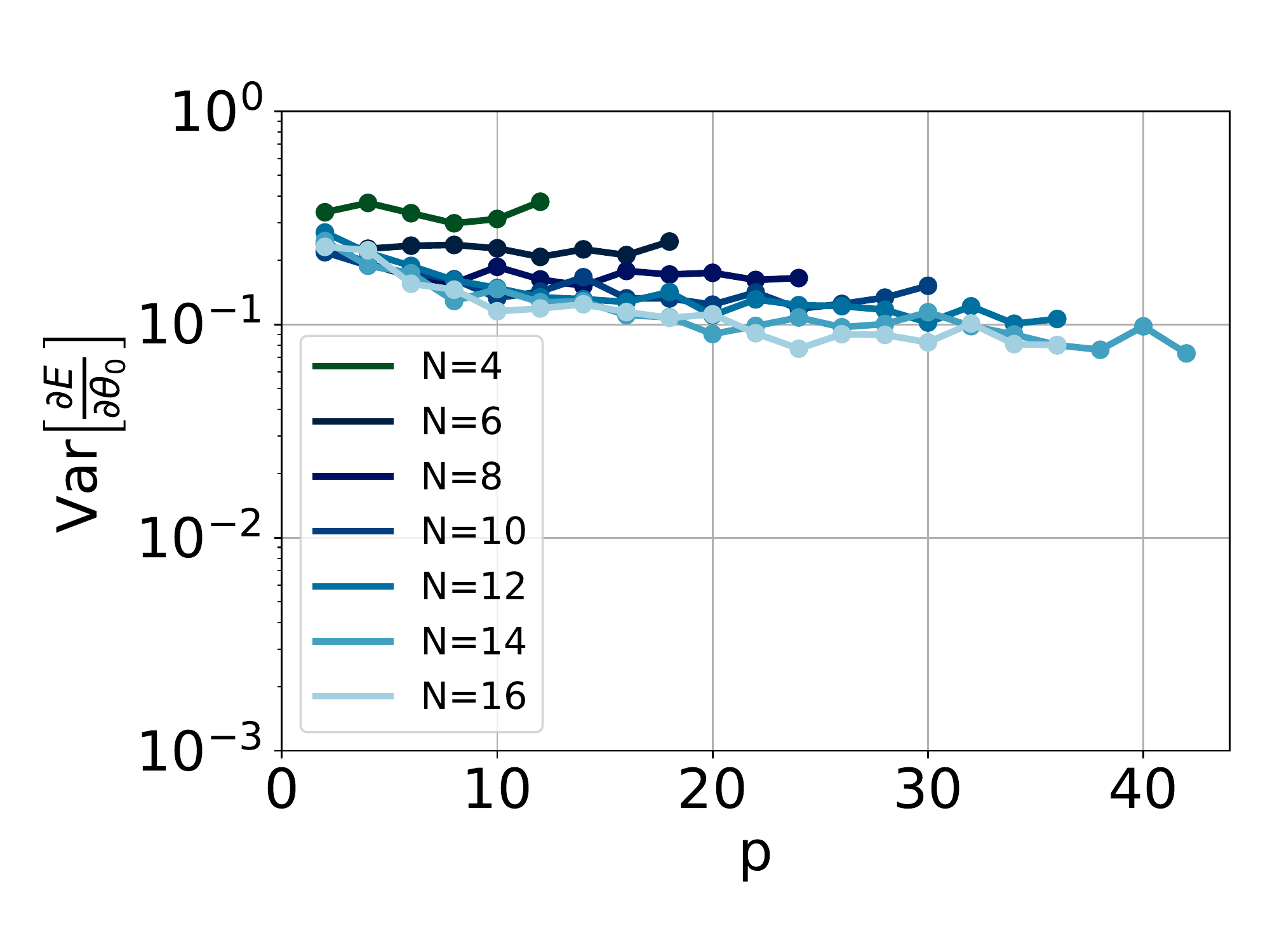} 
}

\subfloat[\label{fig:xxzgrads16}XXZ model]{
    \centering
    \includegraphics[width=0.95\columnwidth]{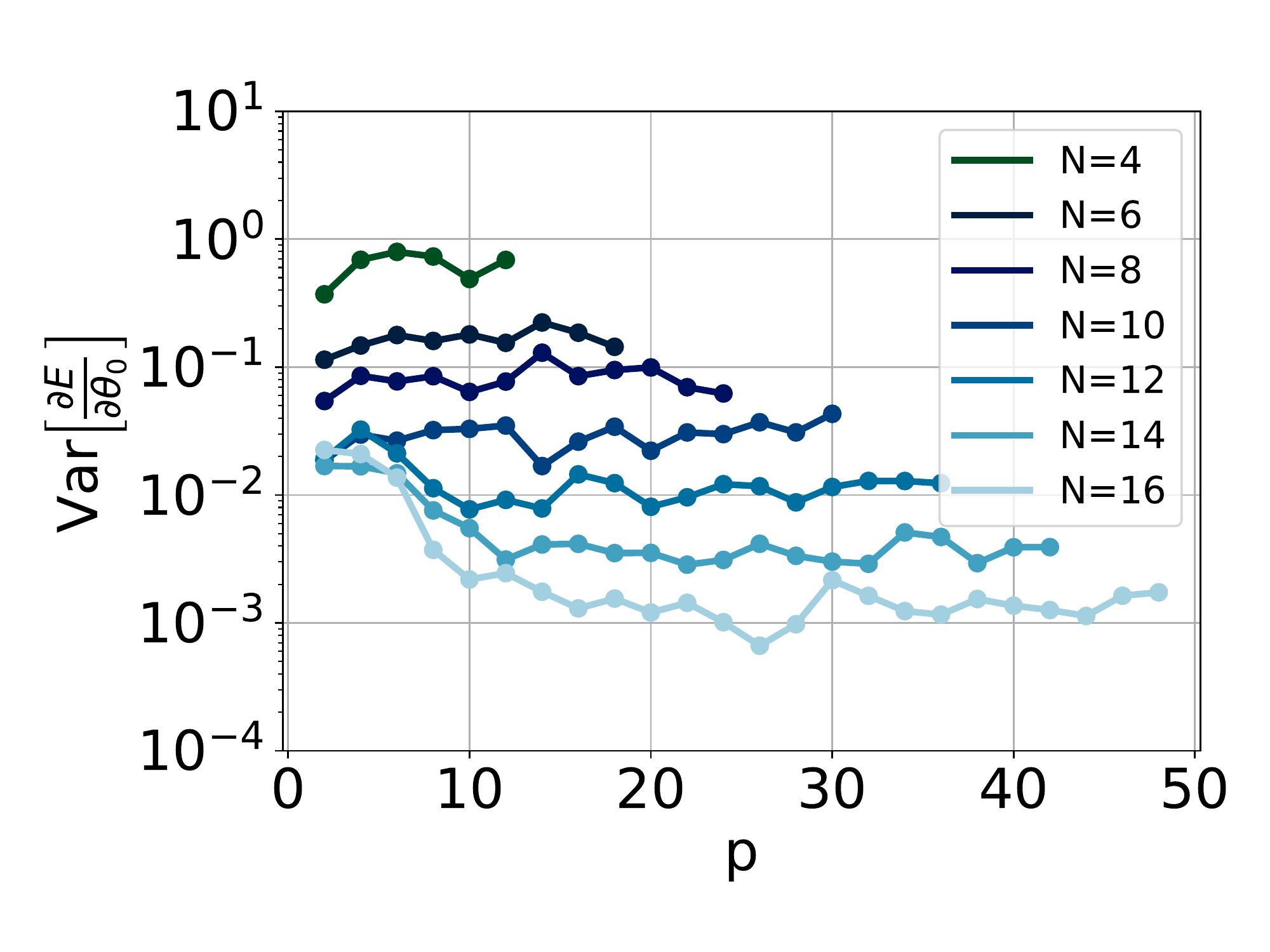}
}
\caption{Variance of the gradients of a single $Z_1 Z_2$ term with respect to $\theta_0$ as a function of the number of qubits at initialization. The number of samples used per $N$ for each $p$ is 20. (a) For the TFIM, the gradient variance decay is almost constant for all $N$. (b) The XXZ model gradient variance is still exponential, although the effect is not as pronounced as for the RQXCs of~\cite{McClean2018barren}, where the $N=16$ variance is two orders of magnitude smaller.}
\label{fig:grads}
\end{figure}

Nonetheless, we determined that the barren plateau problem is significantly ameliorated in the TFIM and mild in the XXZ model. In \cref{fig:grads}, we calculated the variance of gradients as a function of qubits number $N$ and depth $p$ over $20$ random points per $N$ and per $p$. For the TFIM, the flatness of the variance curve indicates no barren plateau problem. However for the XXZ model, we see an exponential decay, but this decay is not as strong as in RQCs~\cite{McClean2018barren}. The scaling of the mean gradient magnitudes follows a similar pattern. Nonetheless, we can reliably find an accurate solution when choosing an identity start (see appendix \cref{app:numres}), where the barren plateau problem is absent. Indeed, sampling gradients close to the identity initialization gives a constant gradient variance for all $N$. This indicates that the vanishing gradient problem can be circumvented by choosing a suitable initialization strategy.

\subsection{The entangling power of HVA circuits}\label{sec:entpower}

For a 1D gapped quantum system, the entanglement entropy of the ground state obeys an area law~\cite{Hastings_2007_area, Arad_2012_area, arad2013area}, i.e., the entanglement entropy grows proportionally to the boundary area $\abs{\partial I}$ of the subsystem $\rho_A$:
\begin{align*}
    S(\rho_A) = \mathcal{O}(\abs{\partial I}).
\end{align*}
In 1D, the boundary area $\abs{\partial I}$ is either $1$ (for an open chain) or $2$ (for a closed chain), and the area law simply says that the entanglement entropy should be constant as $N$ increases. For a 1D conformally invariant gapless (critical) system, the entanglement entropy of the ground state has a logarithmic scaling instead~\cite{Calabrese_2004}, i.e.,
\begin{align*}
    S(\rho_A) = \mathcal{O}(\log(n)).
\end{align*}
Entangling power is an important factor for characterizing the expressiveness and efficiency of many variational ans\"atze in condensed matter physics. It characterizes how much entanglement (measured by the entanglement entropy) can be generated by the variational circuit. For example, in the matrix product state representation, the entangling power is limited by the so-called \emph{bond dimension} $D$ which affects the expressive power and computational cost of the ansatz. For a 1D gapped system with energy gap $\epsilon$, the ground state can be approximated well by an MPS with sublinear bond dimension $D = \exp(\Tilde{O}(\frac{\log^{3/4}n}{\epsilon^{1/4}}))$~\cite{gharibian2015quantum}. In the case of HVA, the amount of entanglement generated by the circuit depends on the depth $p$ of the circuit. Indeed, we observed in \cref{fig:mp} numerically that the HVA circuits for the TFIM and XXZ model have enough entangling power to express the ground states. 

\begin{figure}[htb!]
    \centering
    \includegraphics[width=0.9\columnwidth]{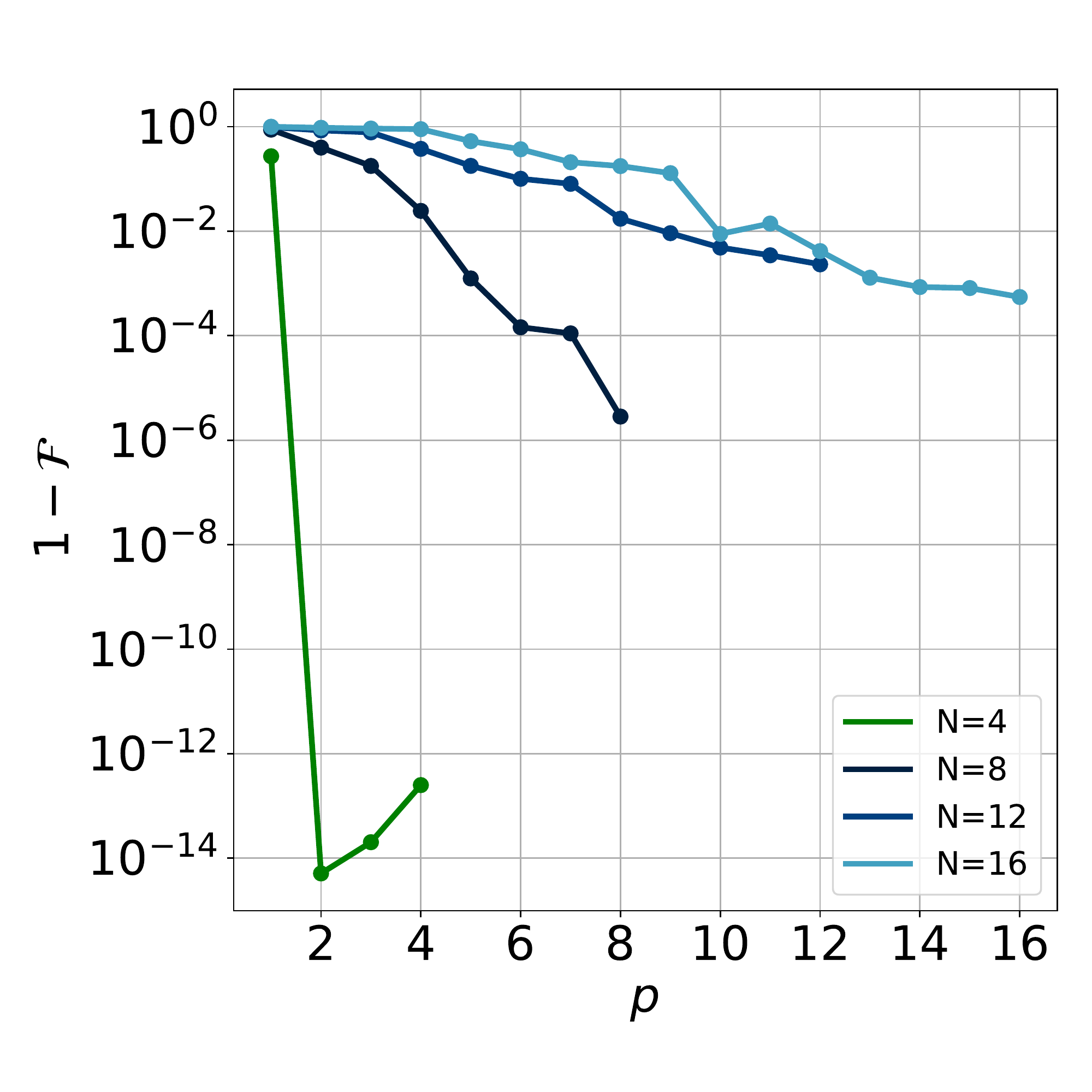}
    \caption{Infidelities found after optimization for the MHS Hamiltonian. The circuit is initialized with an identity start. For the $4$-qubit case we get close to machine precision, and hence the fidelities are unstable.}
    \label{fig:mhs_performance}
\end{figure}

As a demonstration that the full entangling power of HVA can be utilized effectively, we solve for the ground state of the so-called modified Haldane-Shastry (MHS) Hamiltonian. This model has long range interactions and is expected to have power-law entanglement scaling in the ground state~\cite{Haldane1988mhs, Shastry1988mhs}. The MHS Hamiltonian is given by 
    \begin{align*}
        H_{\text{MHS}} = \sum_{j<k}^N \frac{1}{d^2_{jk}}(-\sigma_x^j\sigma_x^k -\sigma_y^j\sigma_y^k +\sigma_z^j\sigma_z^k),
    \end{align*}
where $d_{jk} = \frac{N}{\pi}\abs{\sin(\pi(j-k)/N)}$. Due to the form of the Hamiltonian, we can use the same HVA \eqref{eq:qaoa_xxz} as for the XXZ model. In \cref{fig:mhs_performance} we see that it is possible to find the ground state with $>99.7\%$ fidelity using a depth $p=N$ circuit for $N=4,8,12,16$. 

\section{Conclusion}\label{sec:conclusion}


In this work, we shed light on some of the desirable properties of HVA as a critical ingredient in the variational quantum eigensolver algorithm. In particular, we show evidence that there are only mild or entirely absent barren plateaus in HVA. This is strikingly different from the commonly used random quantum circuits. Moreover, we also observe an over-parameterization phenomenon in HVA. Similar to what was observed in the deep neural networks, the optimization landscape of HVA becomes increasingly better as the ansatz is over-parameterized and eventually becomes trap free as the over-parameterization reaches a certain threshold. In contrast with the case of learning Haar random unitaries, we observe that such threshold in HVA scales at most polynomially with the system size. Finally, we provide numerical evidence that HVA can be used to find the ground state of the MHS Hamiltonian, which has a power-law scaling entanglement. We believe that our findings point to important features of HVA that will lead to a deeper understanding of its effectiveness, and point the way to developing more sophisticated ans\"atze for other many-body problems, as well as more informed optimization/initialization strategies.

As for future work, since most 1D quantum many-body systems can be simulated efficiently with classical methods, the crucible for HVA will be 2D systems. If low-depth circuits are capable of reproducing non-trivial 2D quantum states, then one can start thinking when a quantum advantage can be reached for systems where classical methods are computationally expensive or even ineffective. The effectiveness of the identity initialization, both in terms of the absence of vanishing gradients and reliability of finding a good ground state approximation is striking. Scrutinizing the mechanism for why this is the case will require a deeper understanding of the energy landscape of HVA. Our preliminary results for the XXZ model and TFIM on rectangular lattices show that this initialization strategy remains effective even for 2D systems.

Lastly, the over-parameterized regime is a double edged sword. On the one hand, it implies that we can improve the energy landscape by increasing the depth of the circuit, ameliorating the effects of local minima. On the other hand, the growth in circuit depth, may well nullify this increase in performance due to the longer coherence times required and multiplicative gate errors. In order to asses how useful this regime is for hardware implementations, it would require an understanding of the effect that noise has on the optimization in the over-parameterized regime. The recent work \cite{wang2020noiseinduced} of Wang~et~al. indicates that for a class of VQE ans\"atze including the quantum alternating operator ansatz, there could be severe noise-induced barren plateaus when the number of layers scales polynomially. However, for the practical performance of general HVA, a more careful analysis of the trade-off between the benefits of over-parameterization and the detrimental effects of noise-induced barren plateaus is needed. Moreover, research on designing more effective variational quantum circuits based on HVA should also be pursued.


\section{Acknowledgements}
H.Y. is supported by NSERC Discovery Grant 2019-06636 and a Google Quantum Research Award. Y.B.K. is supported by the NSERC of Canada and the Killam Research Fellowship from the Canada Council of the Arts. J.C. acknowledges support from NSERC, the Shared Hierarchical Academic Research Computing Network (SHARCNET), Compute Canada, Google Quantum Research Award, and the CIFAR AI chair program. Resources used in preparing this research were provided, in part, by the Province of Ontario, the Government of Canada through CIFAR, and companies sponsoring the Vector Institute \url{www.vectorinstitute.ai/#partners}. C.Z. acknowledges partial support from the Postgraduate Affiliate Award from the Vector Institute. 

\bibliographystyle{unsrt}
\bibliography{library.bib}

\begin{appendices}

\section{Computational Details \label{app:compdet}}
For the implementation of our quantum circuits, we use \texttt{Zyglrox}~\cite{Wiersema2020zyglrox}, a powerful TensorFlow-based quantum simulator. For the classical optimization process, we use Adam  (adaptive moment estimation)~\cite{Kingma2014}, a gradient descent-based optimizer, which is widely used in the machine learning community. Compared to vanilla gradient descent and its other variants, Adam updates the learning rates adaptively on a per parameter basis by using estimates of the first and second moments of the gradients. In our own investigation for solving the ground energy problem with HVA, Adam outperformed all the other optimizers available in TensorFlow, with respect to fidelity and convergence times.

Unless stated otherwise, the stopping criterion for our optimization is defined as $\abs{E(\bm{\theta_t}) - E(\bm{\theta_{t+1}})} < \num{1e-13}$ where $t$ is the iteration number. The maximum number of iterations is set to 15000. We use an initial learning rate $r=0.01$ for Adam which gives reasonably consistent results across all the models. Through our own investigation into initial Adam learning rates, we found a learning rate $ \num{1e-3} \leq r \leq \num{4e-2}$ to be a good choice for the optimization for both the TFIM and the XXZ models, as it balances optimization accuracy and convergence speed.

These simulations were performed on the University of Toronto Computer Science Department servers, which housed either AMD Ryzen Threadripper 2990WX, or Silicon Mechanics Rackform iServ R331.v4 with two 12-core Intel E5-2697v2 CPUs with access to at most 10gb of RAM. More computationally expensive simulations were done using Nvidia GeForce GTX 1050Ti GPUs.

On the Vector cluster we used Intel(R) Xeon(R) Silver 4110 CPUs and Nvidia 480 T4 GPUs with access to at most 32gb of RAM.

\vspace{-2mm}
\section{Additional Numerical Results\label{app:numres}}
\begin{figure*}[htb!]
\centering
\subfloat[\label{fig:tfitogether} TFIM]{
    \centering
    \includegraphics[width=0.45\textwidth]{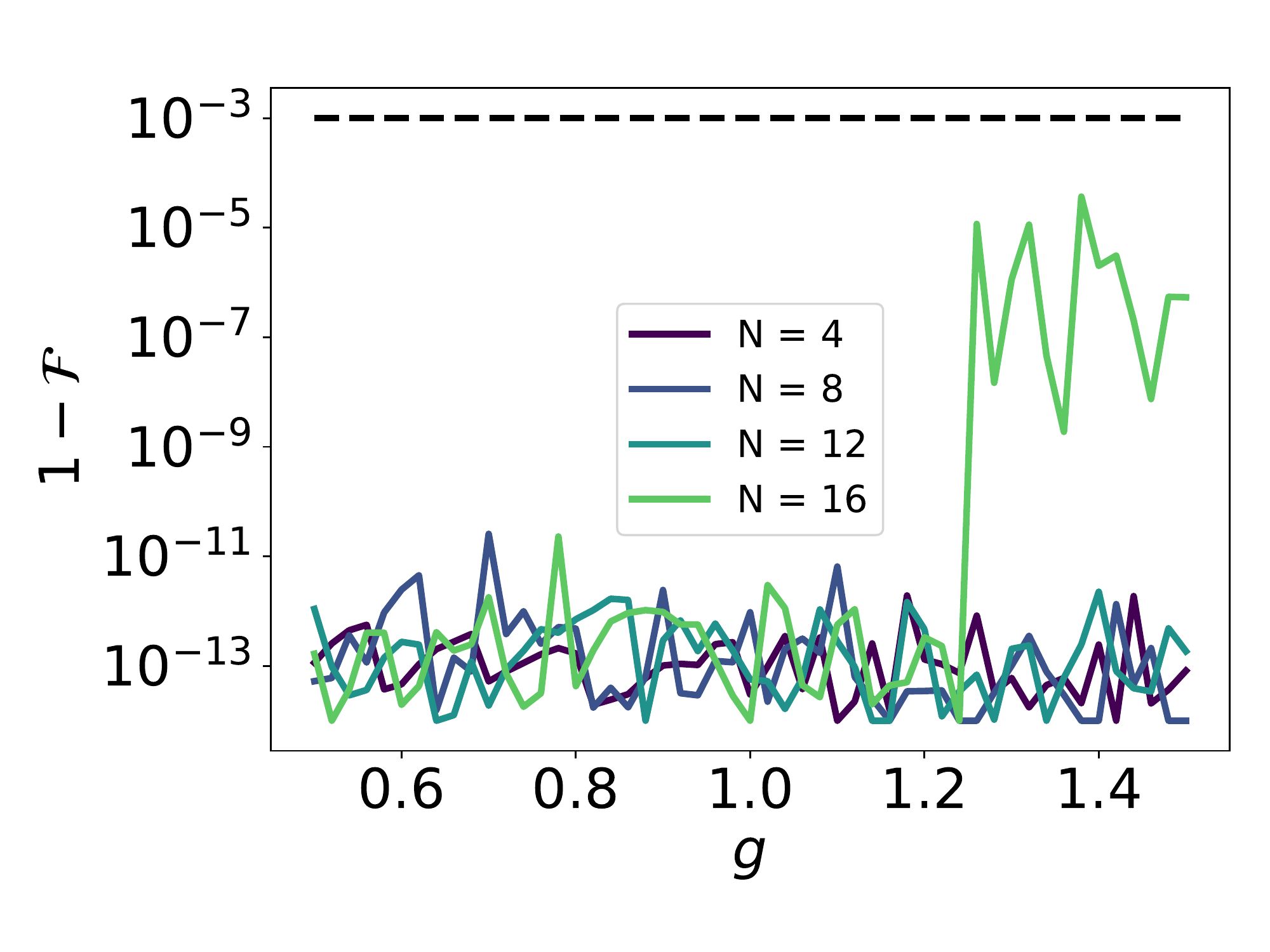}
    }
\subfloat[\label{fig:xxztogether} XXZ model]{
    \centering
    \includegraphics[width=0.45\textwidth]{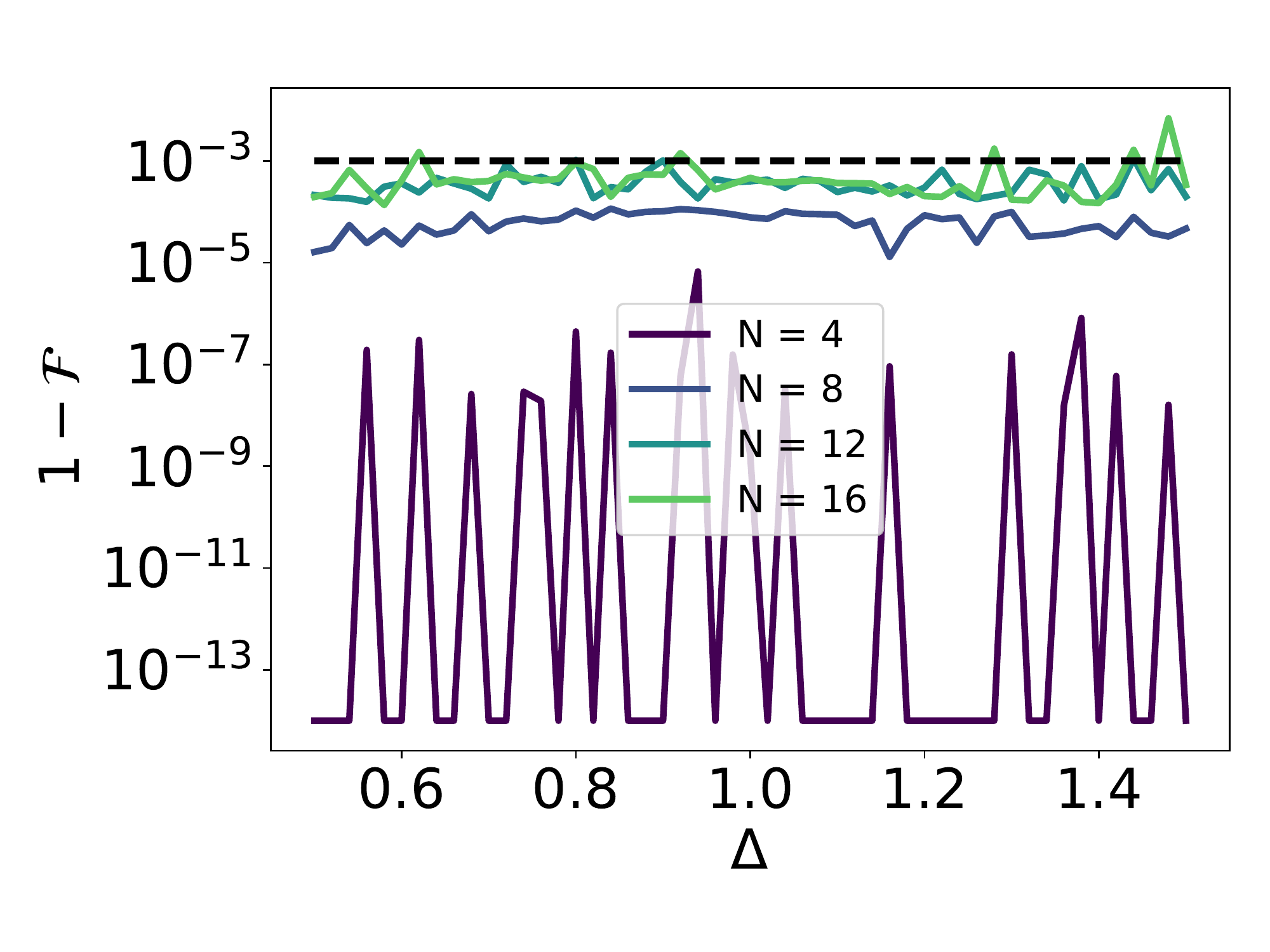}
    }
\caption{Infidelities as a function of the order values $g$ and $\Delta$ for the TFIM and XXZ model respectively. For these results we use the identity initialization combined with a depth $p=N/2$ circuit. The dashed black line indicates the cutoff for $99.9\%$ fidelity. (a) For the TFIM we can obtain machine precision results, except in the region $g>1.24$ for $N=16$. In this region, the optimization has not fully converged, but is stopped after $15000$ iterations. We note that for increasing $N$ the time until convergence is polynomial in $N$ (not shown here), similar to what was observed in~\cite{wierichs2020avoiding}. Additionally, we observe a worsening of this scaling with increased $g$. (b) For the XXZ model we are unable to consistently reach machine precision fidelities. In addition, the fidelities we find become worse as $N$ increases. However, except for a couple of outliers we are able to get $>99.9\%$ fidelities for $N\leq16$ or all order values.}
\label{fig:performance}
\end{figure*}

In \cref{fig:performance} we show the infidelities we find after optimization for $g\in\{0.5, 0.52,\ldots, 1.5\}$ and $\Delta \in \{0.5, 0.52,\ldots, 1.5\}$ for the TFIM and XXZ model, respectively. Our numerical results for the aforementioned ranges of order values are available in the dataset module of \texttt{Tensorflow Quantum}~\cite{Broughton2020tfq}.

\begin{figure*}[htb!]
\centering
\subfloat[\label{fig:tfioverparam_min} TFIM]{
    \centering
    \includegraphics[width=0.9\columnwidth]{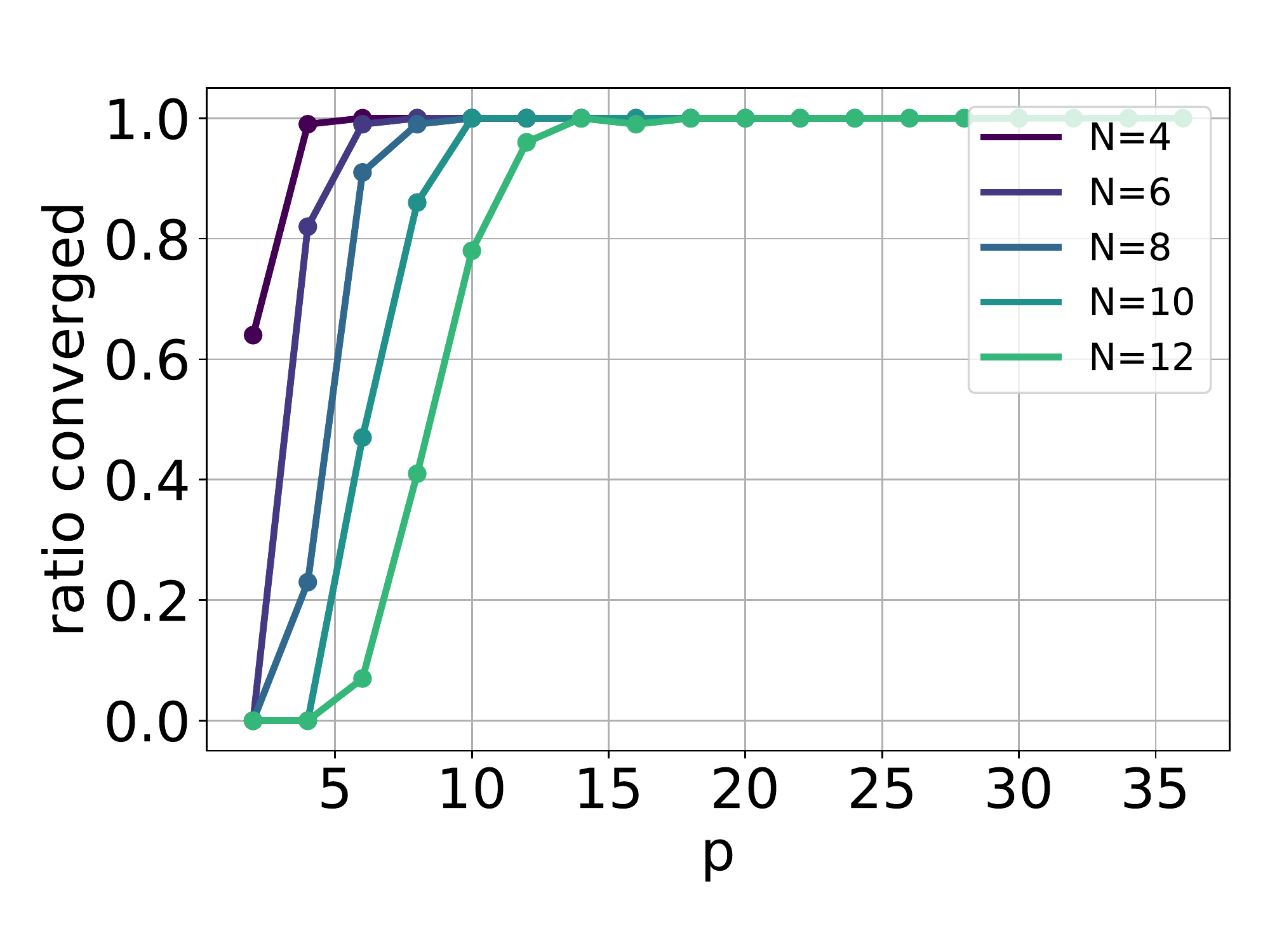}
}
\subfloat[\label{fig:xxzoverparam_min} XXZ model]{
    \centering
\includegraphics[width=0.9\columnwidth]{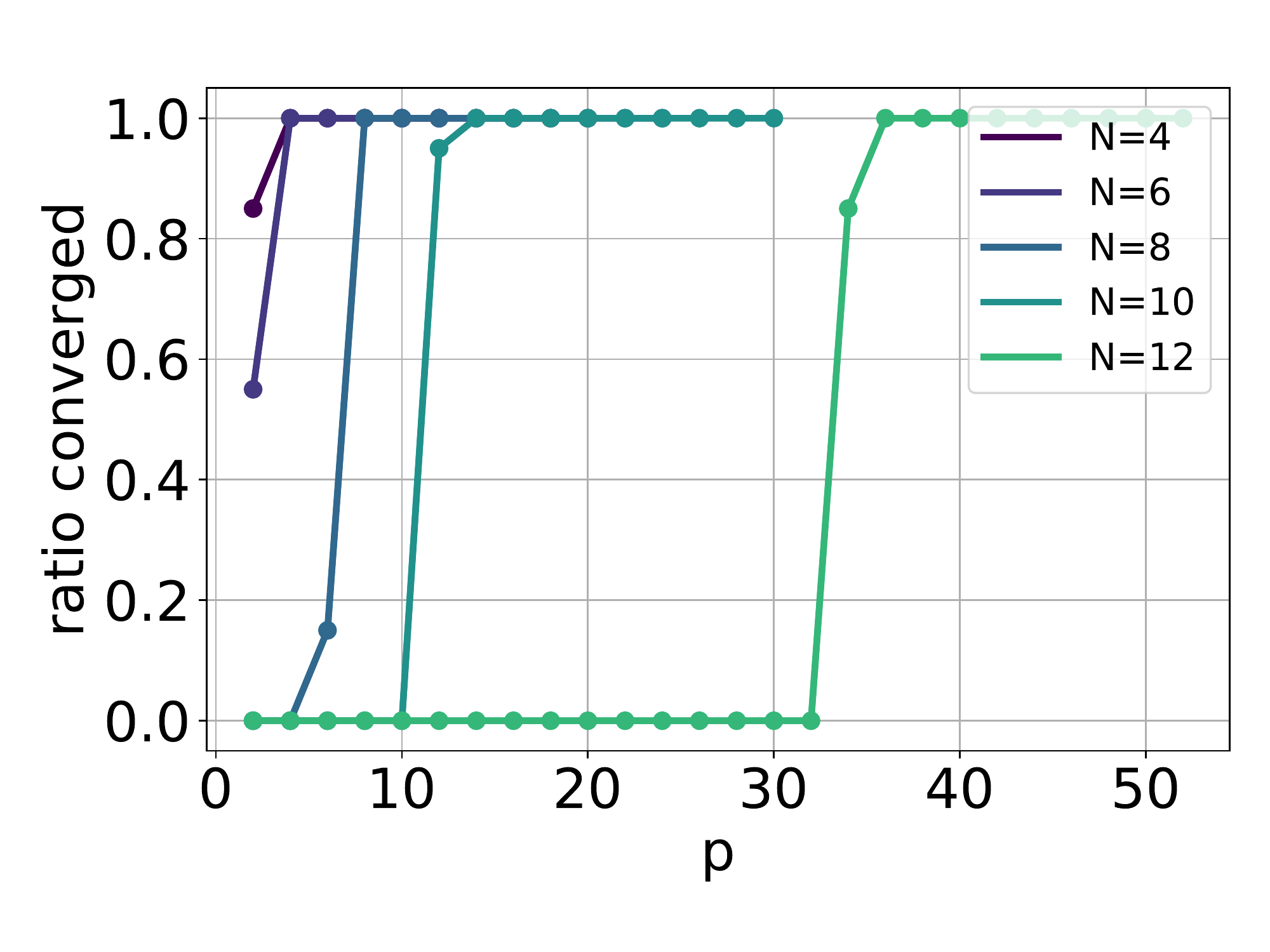}
}
\caption{Ratio of random initializations that converge to the ground state. We consider a run converged if $\epsilon_{\text{res}} \leq 1e-4$.}
\label{fig:percentage}
\end{figure*}

In \cref{fig:percentage}, we show the ratio of converged random initializations as a function of the circuit depth $p$ for different system size $N$.


\section{Dynamics of Entanglement Entropy during Optimization}\label{app:entropydynamics}

\begin{figure*}[htb!]
    \centering
    \subfloat[\label{fig:tfi_random_ent} TFIM $p=4$]{
        \centering
        \includegraphics[width=0.4\textwidth]{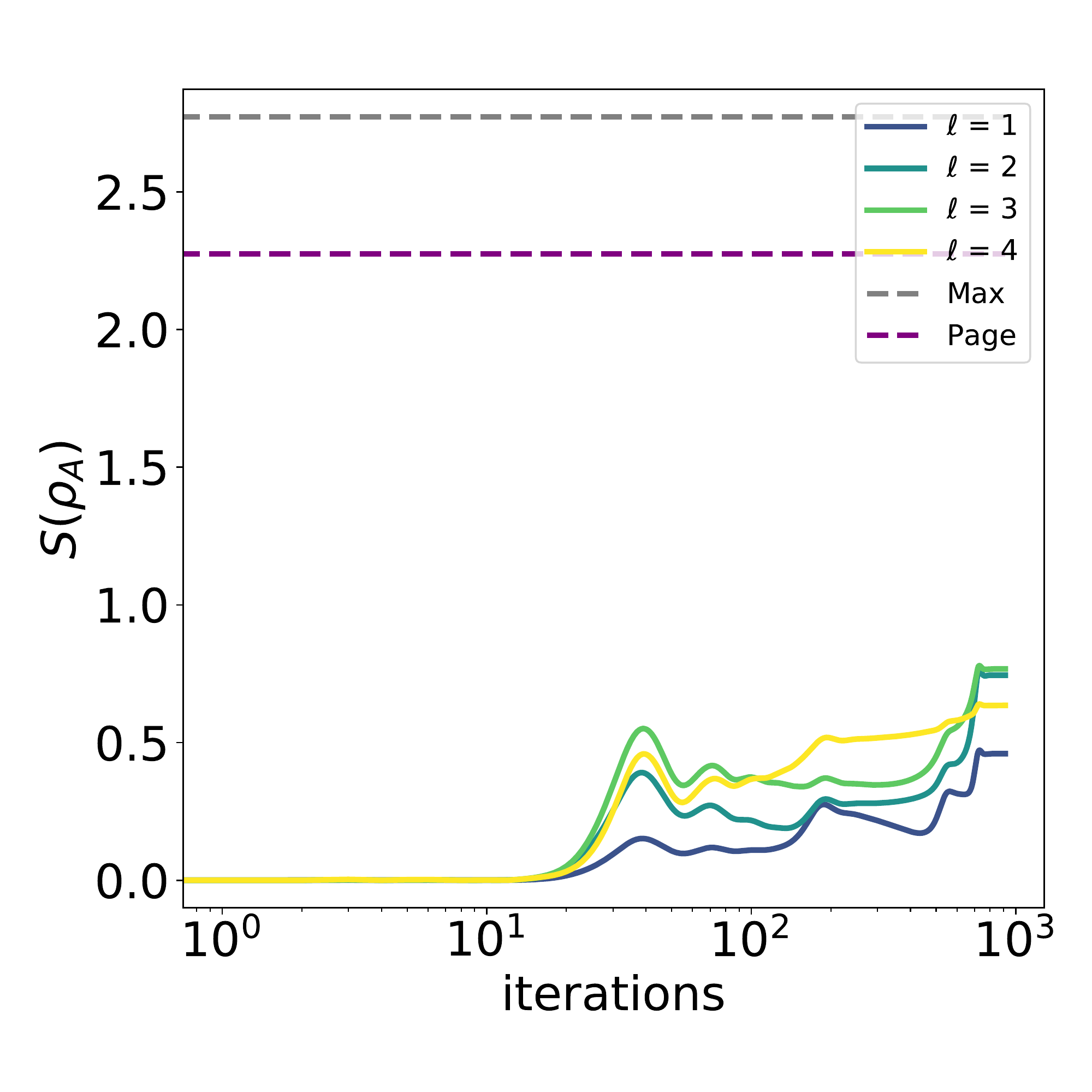}
        }
    \subfloat[\label{fig:tfi_random_ent_over} XXZ model $p=8$]{
        \centering
        \includegraphics[width=0.4\textwidth]{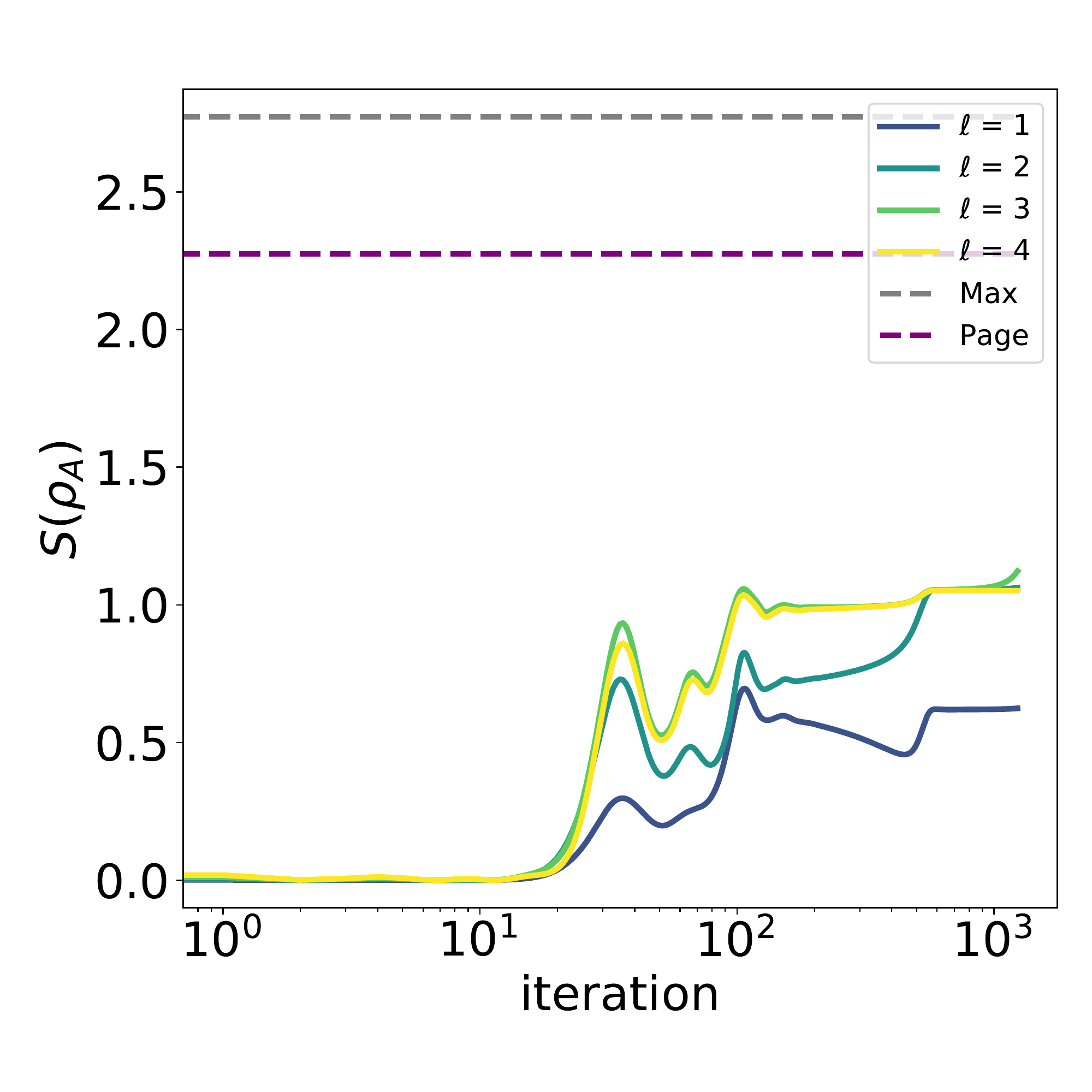}
        }
    
    \subfloat[\label{fig:xxz_random_ent} TFIM $p=4$]{
        \centering
        \includegraphics[width=0.4\textwidth]{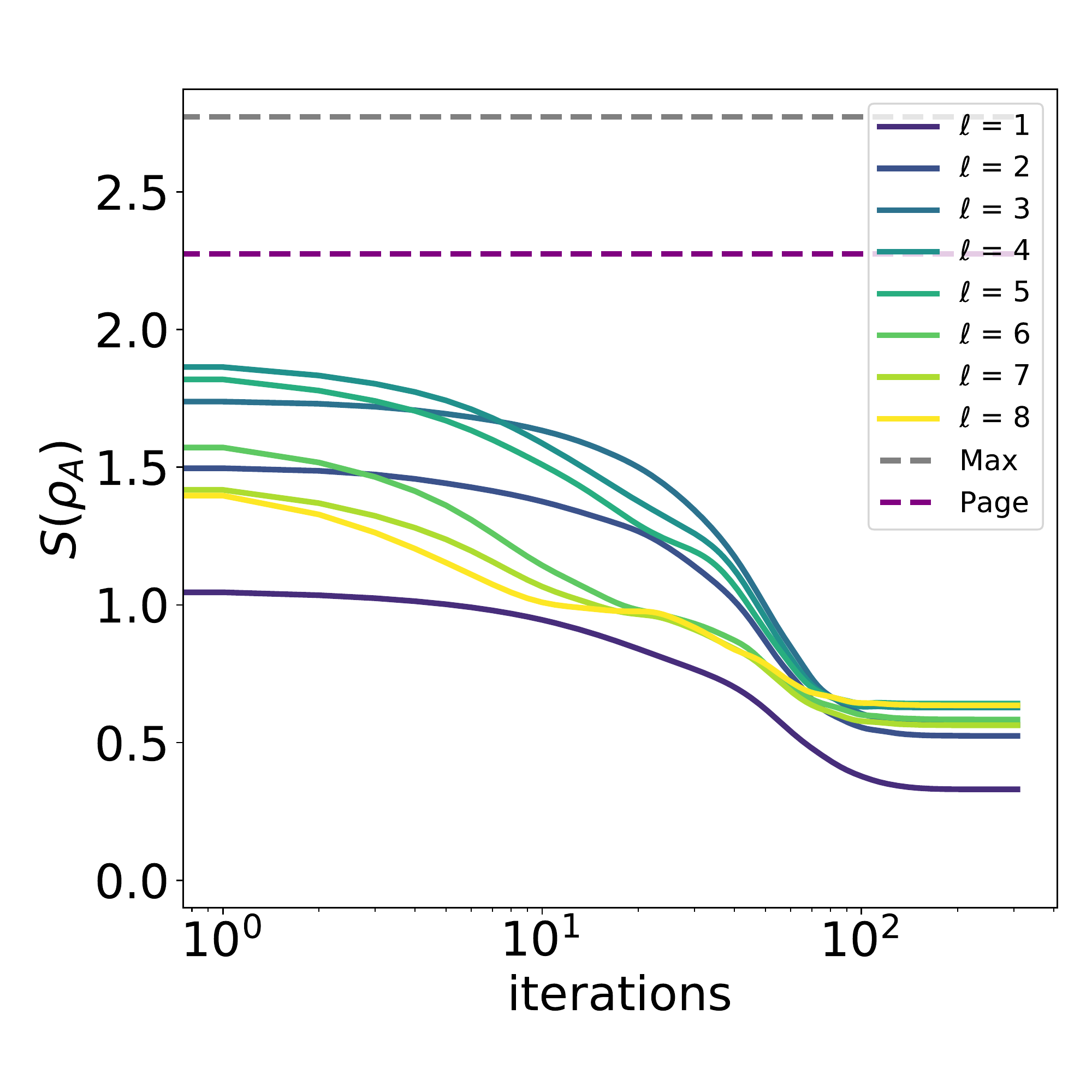}
        }
    \subfloat[\label{fig:xxzent8_over} XXZ model $p=8$]{
      \centering
        \includegraphics[width=0.4\textwidth]{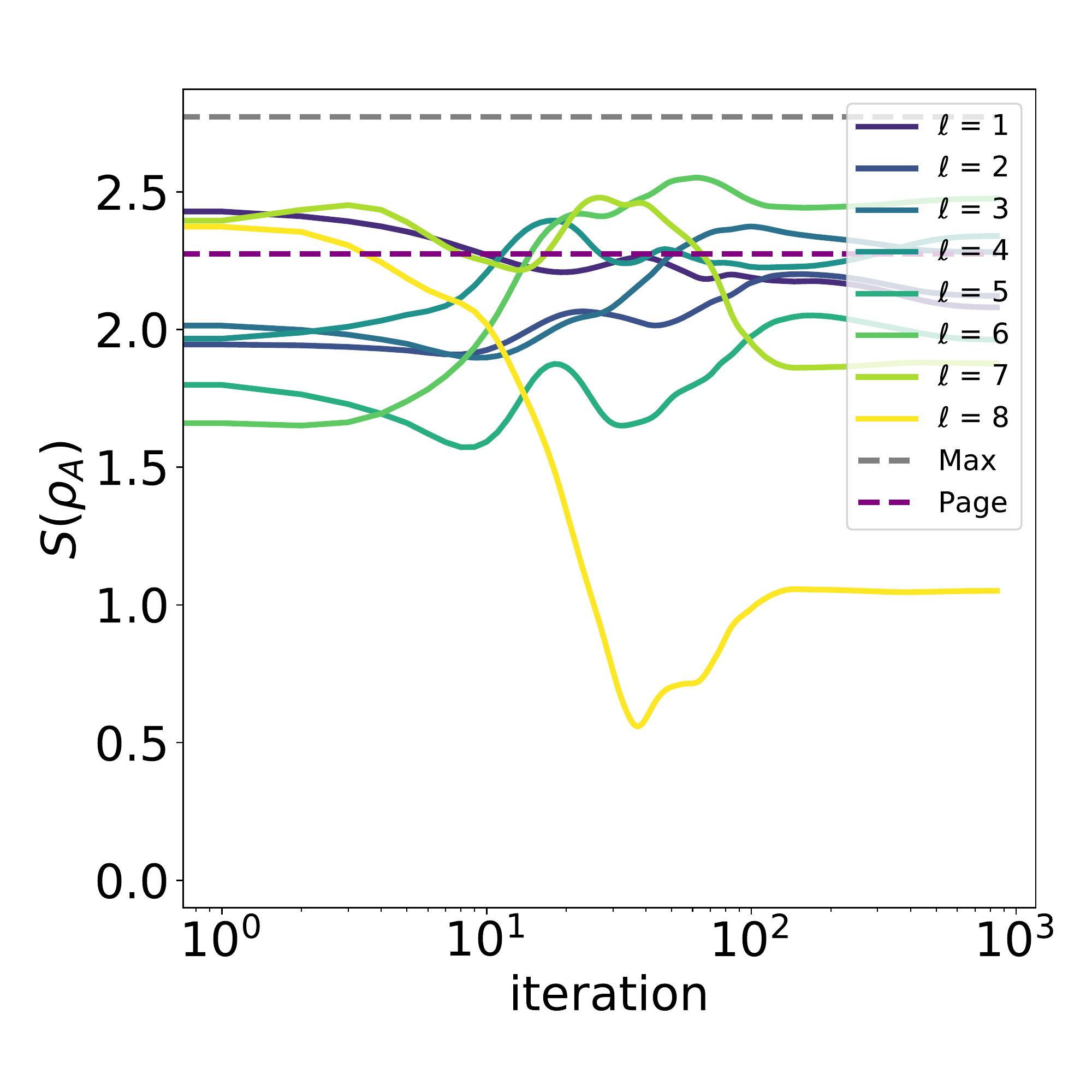}
        }
    \caption{Dynamics of entanglement entropy at each layer during optimization. Each separate line indicates the entanglement entropy of the state in layer $l$. The gray dashed line denotes the maximum possible entanglement and the purple line gives the Page entropy. For all figures, the final state is a $>99.9\%$ fidelity state. (a) Identity  initialization for an $8$-qubit TFIM with $g=1.0$ and $p=4$. (b) Same TFIM with a random-state initialization and over-parameterization $p=8$. (c) Random-state initialization for an $8$-qubit XXZ model with $\Delta=1.0$ and $p=4$. (d) Typical XXZ model dynamics for a random-state initialization and over-parameterization $p=8$.}
    \label{fig:dynamics_entanglement}
\end{figure*}

\begin{figure*}[htb!]
\centering
    \subfloat[\label{fig:tfi_final_state_1} TFIM $g=1.0$]{
    \centering
    \includegraphics[width=0.33\textwidth]{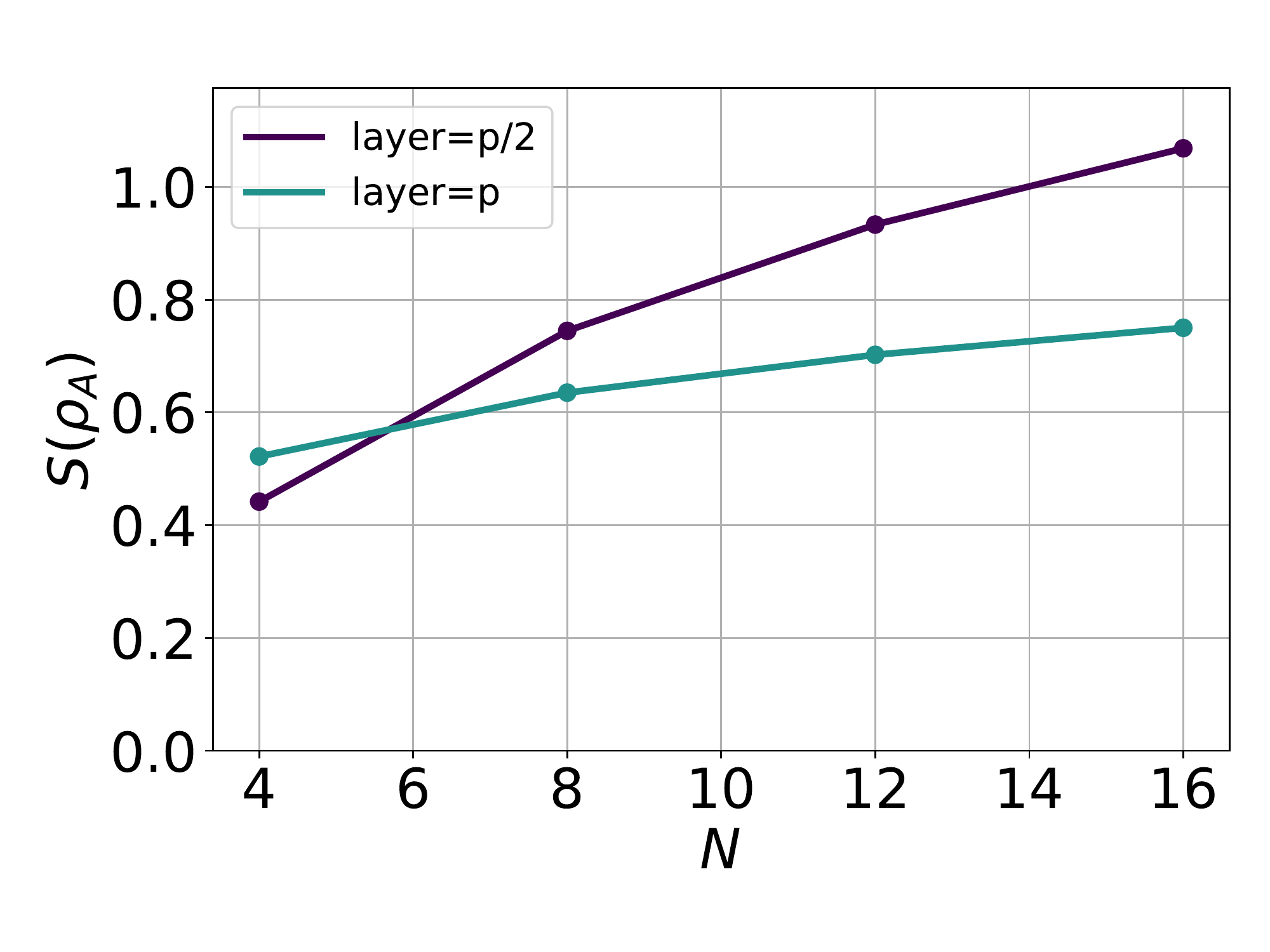}
    }
    \subfloat[\label{fig:tfi_final_state_0.5} TFIM $g=0.5$]{
      \centering
        \includegraphics[width=0.33\textwidth]{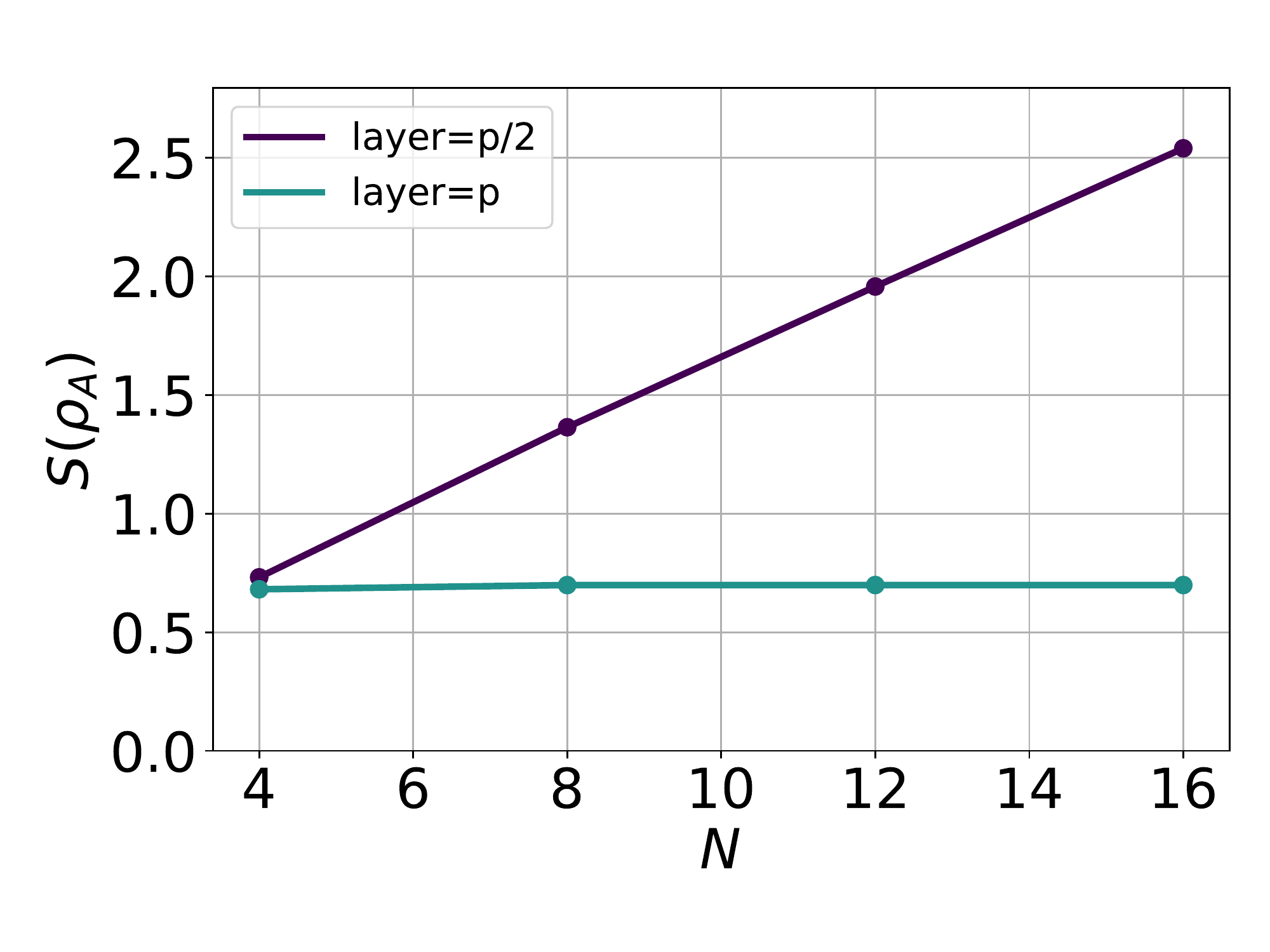}
    }
    \subfloat[\label{fig:xxz_final_state_1.0} XXZ model $\Delta=1.0$]{
        \centering
        \includegraphics[width=0.33\textwidth]{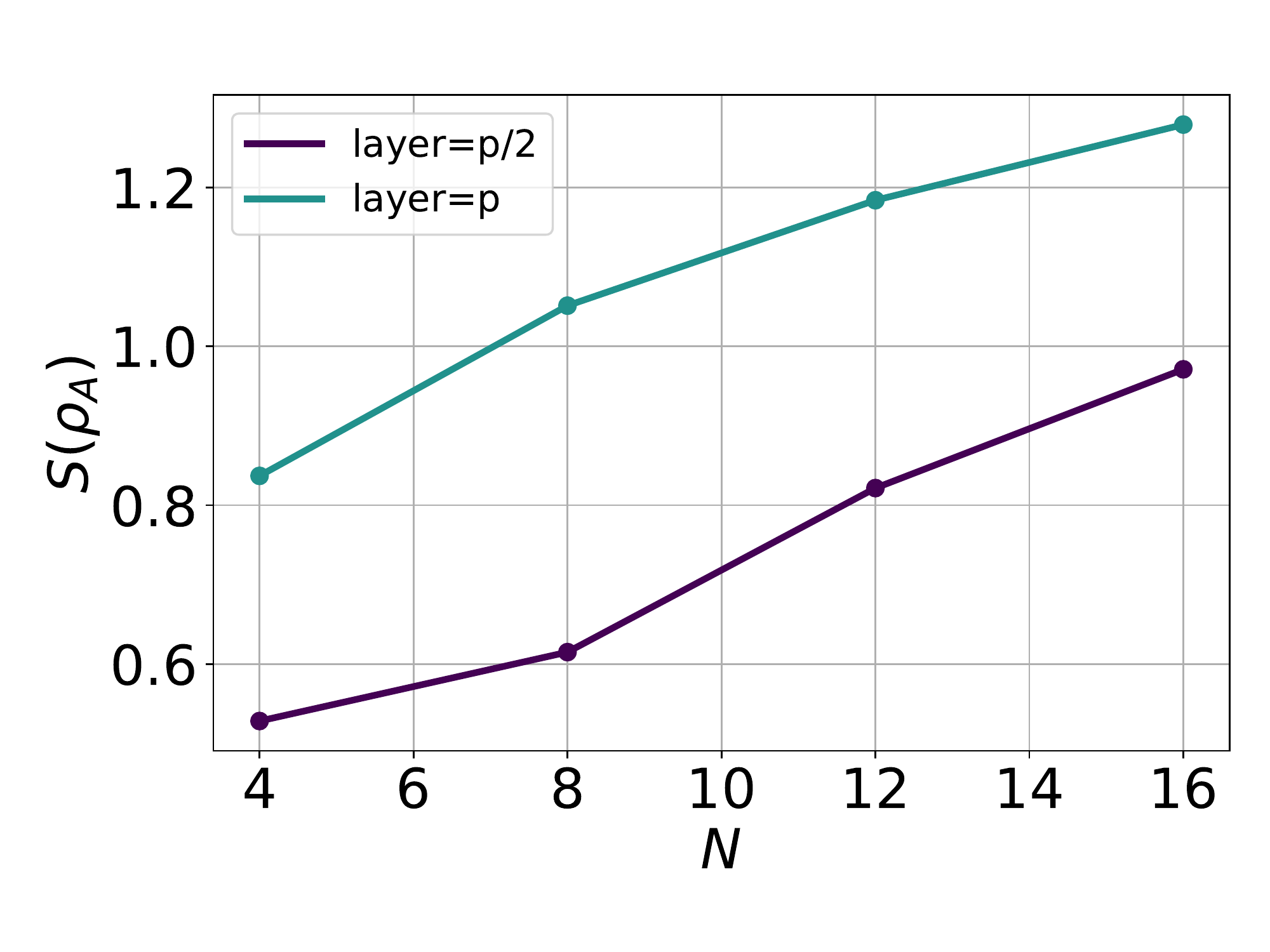}
    }
    
    \caption{Scaling of the entanglement entropy of the converged state after $p/2$ and $p$ layers. (a) For the TFIM at the critical point, the ground state entanglement entropy has a logarithmic correction with increasing $N$. The entanglement halfway through the circuit is larger than in the final layer. (b) For a non-critical point, the ground state entanglement entropy is constant, but the entanglement entropy halfway through the circuit scales linearly with system size. (c) For the XXZ model, in addition to the logarithmic scaling of the entanglement entropy, the final layer entanglement is consistently higher than in the $p/2$ depth layer.}
    \label{fig:final_state}
\end{figure*}

To further elucidate the difference in initialization strategies we qualitatively study the dynamics of the entanglement entropy during optimization. In \cref{fig:dynamics_entanglement}, we calculate the entanglement entropy of $\rho_A$ at each layer of the circuit during the optimization. Although not much can be said about the intermediate states for the random state initialization, except that they are highly entangled, the entanglement entropy dynamics for the identity initialization have a distinct structure that is consistent as we increase the system size. In \cref{fig:final_state}, we compare the scaling of the entanglement entropy for the identity start halfway through the circuit for different system sizes.

\end{appendices}

\end{document}